\documentclass[12pt]{article}
\usepackage[margin=1.1in]{geometry} 
\usepackage{amsmath}
\usepackage{times}
\usepackage{graphicx}
\usepackage{color}
\usepackage{multirow}
\usepackage[authoryear]{natbib}
\usepackage{rotating}
\usepackage{bbm}
\usepackage{latexsym}

\usepackage{amsmath}
\usepackage{amssymb}
\usepackage{mathrsfs}
\usepackage{array}
\usepackage{float}
\usepackage{soul}
\usepackage{setspace}
\usepackage{comment}
\usepackage[normalem]{ulem}
\usepackage{caption}
\usepackage{pdfpages}

\newenvironment{proposition}[2][]{%
  \noindent \textbf{Proposition~#2.} \itshape}{\par}

\newenvironment{proof}{{\noindent\bfseries Proof.}}{\hfill$\square$\par}

\begin{document}

\begin{center} {\LARGE \textbf{Quantifying information stored in synaptic connections rather than in firing activities of neural networks}}
 \end{center}

\ \\
\begin{center}
{\bf Xinhao Fan$^{\displaystyle 1, \displaystyle *}$, Shreesh P Mysore $^{\displaystyle 2, \displaystyle 1,\displaystyle 3, \displaystyle *}$}\\
{$^{\displaystyle 1}$ The Solomon H. Snyder Department of Neuroscience, Johns Hopkins University, Baltimore, MD, USA.}\\
{$^{\displaystyle 2}$ Department of Psychological and Brain Sciences, Johns Hopkins University, Baltimore, MD, USA.}\\
{$^{\displaystyle 3}$ Kavli Neuroscience Discovery Institute, Johns Hopkins University, Baltimore, MD, USA.}\\
{$^{\displaystyle *}$ Corresponding author. (xfan20@jhu.edu, mysore@jhu.edu)}\\
\end{center}
%

\thispagestyle{empty}
\markboth{}{NC instructions}
\ \vspace{-0mm}\\
%
\begin{center} {\bf Summary } \end{center}

A cornerstone of our understanding of both biological and artificial neural networks is that they store information in the strengths of synaptic connections among the neurons. However, in contrast to the well-established theory for quantifying information encoded by the firing activity of neural networks, there does not exist a framework for quantifying information stored in the network's connection distribution itself. Here, we develop a theoretical framework for synaptic information by using densely connected Hebbian networks performing autoassociative memory tasks and by modeling data patterns to be stored as log-normal distributions.
Specifically, we  derive analytical approximations for Shannon mutual information between the data and singletons, pairs, and arbitrary $n$-tuples of synaptic connections within the network. Our framework corroborates well-established insights regarding pattern storage capacity, supports the principle of distributed coding in neural firing activities, and formalizes the heterogeneity inherent in information encoding across synapses in a network. Notably, it discovers synergistic interactions among synapses, revealing that the information encoded jointly by all the synapses exceeds the `sum of its parts'. Taken together, this study introduces a powerful, interpretable framework for quantitatively understanding information storage in the synapses of neural networks, one that illustrates the duality of synaptic connectivity and neural population activity in learning and memory.

\section{Introduction}

The study of neural networks through the lens of information theory has a long and rich history \citep{borst1999information, quian2009extracting, dimitrov2011information, timme2018tutorial}. The analyses therein have focused typically on measuring information encoded by the firing activities of neurons, both in biological \citep{bialek1992reliability, palmer2015predictive} and artificial neural networks \citep{linsker1988self, tishby2015deep}. By contrast, the study of neural networks through the lens of learning and memory posits that information is inherently stored in the synaptic connections, with neural activity patterns serving as a manifestation of the underlying changes in connectivity. This concept was notably articulated by Hebb \citep{hebb-organization-of-behavior-1949}, in his seminal theory of memory and learning, where he described a progression from synaptic modifications to the formation of cell assemblies, and eventually to the development of ``phase sequences" representing sequential activation of assemblies. Building on this perspective, subsequent research has demonstrated that connectivity encodes essential information about the world through mechanisms such as changes in synaptic strength \citep{lamprecht2004structural}, the formation of wiring motifs \citep{sporns2004motifs, battiston2017multilayer}, and the emergence of synaptic ensemble patterns in learning \citep{baeg2007learning, bassett2011dynamic} and working memory \citep{mongillo2008synaptic, stokes2015activity, panichello2024intermittent}.

The theoretical analysis of information stored in the distribution of synaptic connections, however, remains largely under-explored. A few studies in the field of machine learning have invoked the concept of `information in weights' for the purpose of regularization during the training of neural networks. They have done so, typically, by introducing noise into weights \citep{hinton1993keeping, achille2017emergence}. However, these analyses primarily provide an upper bound on the information stored in connections, which, while useful for practical applications in network training, do not directly address the question of how much information is actually encoded in the connections. Moreover, for the sake of theoretical simplicity, these approaches often assume that individual connections are probabilistically independent, thereby neglecting the inherently collective nature of coding by synaptic ensembles. 

An alternative line of research, focused on associative neural networks such as Hopfield networks and Willshaw models, first estimates the network's overall storage capacity based on patterns retrievable in the neural firing activities, and then derives the `information per synapse' by dividing the capacity by the total number of connections \citep{willshaw1969non, nadal1991associative, palm1996associative, bosch1998information, knoblauch2010memory}. However, this approach continues to rely on firing patterns to infer synaptic coding, thereby still fundamentally adopting a firing activity perspective. Moreover, division by the synapse count oversimplifies the contributions of individual connections - by assuming uniform efficacy across synapses and by failing to account for the heterogeneous roles that different connections may play in information coding. Importantly, the collective coding of information by synaptic \textit{ensembles} remains unaddressed. 

The major hurdle in quantifying the information stored in synaptic weights is the complex and opaque relationship between data and weight distributions in both biological and artificial neural networks. Unlike the more explicit mapping between data and neural firing activities, weight distributions arise implicitly from the network learning process, rather than through an explicit or closed-form transformation of the data distributions. Consequently, a theoretical analysis of information encoded in ensembles of \textit{connections}, as opposed to ensembles of \textit{cells}, has remained a largely open question. 

In this study, we propose a foundational framework for synaptic coding based on dense Hebbian connectivity,  and data patterns assumed to follow a mixture of log-normal distributions (without loss of generality). The accessibility of the weight values of Hebbian networks, and the tractability of the selected distribution in calculating mutual information  allow us to derive analytical approximations for the information encoded by individual as well as arbitrarily sized ensembles of synaptic connections, ranging from pairs and triplets to larger $n$-tuples. These derivations incorporate two approximations related to the log-normal distribution. The resulting expressions validate established insights about distributed coding and storage capacity from the traditional perspective of network firing activities. Additionally, they reveal intriguing new insights about information synergy among synaptic connections in neural networks. Overall, this research fills an important gap in the field by introducing a theoretical framework for characterizing information storage in neural networks, and highlights the dual significance of synaptic connectivity and neural population activity in neural coding.

\section{Analytical framework: Information stored in synaptic connections}
We begin by considering the $N$ data patterns that are to be stored as  independent, $d$-dimensional random variables. We denote them as $ \{\mathbf{x}^1, \dots, \mathbf{x}^N\} $. Each of these random variables is assumed to follow a multivariate log-normal distribution, a formulation that captures the heavy-tailed statistics observed broadly in diverse neuronal populations \citep{buzsaki2014log}. 
Formally, this is expressed as:
\begin{equation}
    \ln(\mathbf{x}^k) \sim \mathcal{N}(\mathbf{\mu}^k, \Sigma^k), \qquad \mathbf{x}^k \perp \mathbf{x}^l, \forall k \neq l
\label{eq: pattern log normal}
\end{equation}
where $\mathbf{\mu}^k$ is a $d$-dimensional vector representing the mean of the $k^{\text{th}}$ pattern in the logarithmic domain, and $\Sigma^k$ is the associated $d \times d$ covariance matrix.

For  our neural network model, we employ the densely connected network with synaptic weights set according to the Hebbian learning rule. The connectivity matrix of such networks is of dyadic form with continuous real values such that each element is determined entirely by the data patterns. Specifically, the synaptic connection between neuron $ i $ and neuron $ j $ is given by: 
\begin{equation}
w_{ij} = \sum_{k=1}^N \mathbf{x}^k_i \mathbf{x}^k_j
\label{eq:hopWts}
\end{equation}

With this foundation in place, we depart from the traditional approach of investigating the mutual information between the continuous firing dynamics of neurons and the data patterns, and focus instead on the Shannon mutual information between the synaptic connectivity among neurons and the data patterns. The existence of a closed-form expression for the weights in the Hebbian network (eqn. \ref{eq:hopWts}) facilitates this endeavor.

{
\begin{center}
\includegraphics[width=0.7\textwidth]{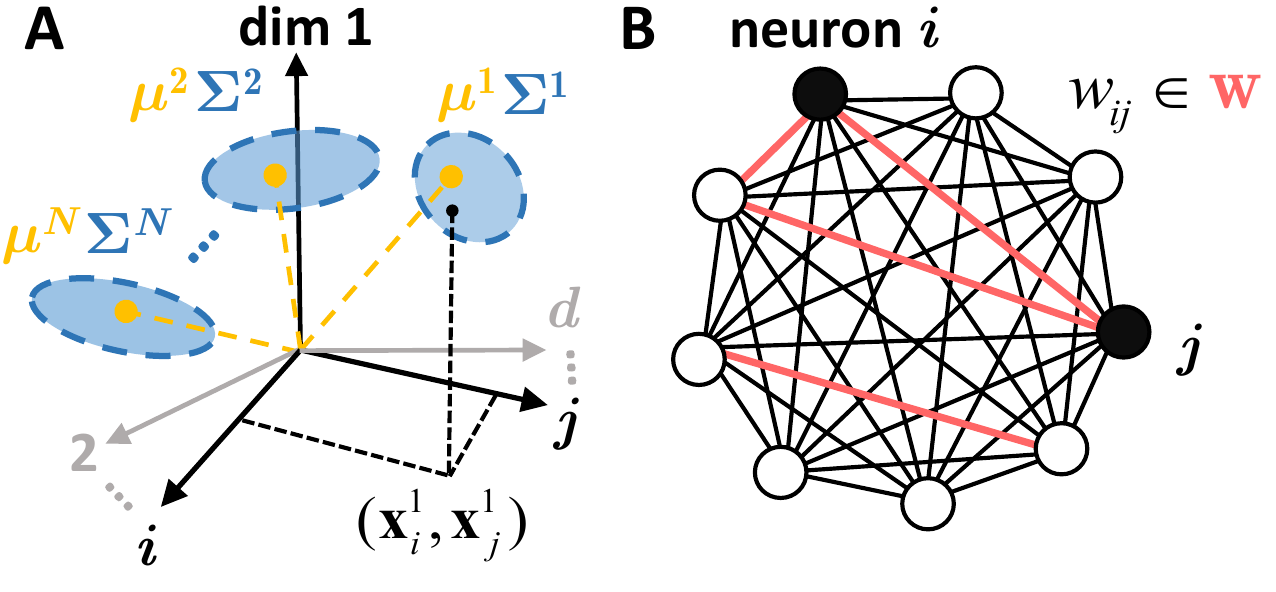}
\end{center}
\captionof{figure}{Model setup. (A) Input data distribution modelled as several independent patterns following multivariate log-normal distribution. (B) Densely connected Hebbian network with an example synaptic ensemble $\mathbf{w}$ (red) composed of four connections. } 
\label{Fig:model}
}

\subsection{Information encoded by one synaptic connection}
Here, we derive the approximation for the mutual information $ MI(w_{ij}; \mathbf{x}^l) $ between a synaptic connection $ w_{ij} $ and a data pattern $ \mathbf{x}^l $, by obtaining closed-form expressions for the two pertinent probability distributions: $ p(w_{ij}) $ and $ p(w_{ij} | \mathbf{x}^l) $. 

For a given pattern $ \mathbf{x}^k $, the vector $ \ln \mathbf{x}^k $ follows a multivariate Gaussian distribution. Consequently, its $ i^{\mathrm{th}} $ component also follows a Gaussian distribution, denoted as $ \ln \mathbf{x}^k_i \sim \mathcal{N}(\mathbf{\mu}^k_i, (\sigma^k_i)^2) $, where the mean $ \mathbf{\mu}^k_i $ is the $ i^{\mathrm{th}} $ component of $ \mathbf{\mu}^k $, and the variance $ (\sigma^k_i)^2 $ is the $ i^{\mathrm{th}} $ diagonal element of the covariance matrix $ \Sigma^k $. Similarly, the sum of the $ i^{\mathrm{th}} $ and $ j^{\mathrm{th}} $ components follows a Gaussian distribution of the form:
\begin{equation}
\ln(\mathbf{x}^k_i) + \ln(\mathbf{x}^k_j) = \ln(\mathbf{x}^k_i\mathbf{x}^k_j) \sim \mathcal{N}(\mathbf{\mu}^k_i + \mathbf{\mu}^k_j, (\sigma^k_i)^2 + (\sigma^k_j)^2 + 2\sigma^k_{ij}),
\end{equation}
where $ \sigma^k_{ij} $ represents the $ (i,j) $ element of the covariance matrix $ \Sigma^k $. Therefore, the random variable $ \mathbf{x}^k_i\mathbf{x}^k_j $ also follows a log-normal distribution. The synaptic connection $ w_{ij} = \sum_{k=1}^N \mathbf{x}^k_i \mathbf{x}^k_j $ is thus the sum of multiple independent log-normally distributed variables.

Fortunately, the problem of approximating the sum of log-normal variables has been studied extensively in  engineering due to the prevalence of such noise distributions in communication systems. A widely used approach in the field, the Fenton-Wilkinson approximation \citep{fenton1960sum}, demonstrates that the sum of log-normal variables can be reasonably well approximated by another log-normal variable using the moment-matching method. We apply this approach to derive the probability distribution $ p(w_{ij}) $; please see Methods (section \ref{sec: proofs}) for proofs of all propositions. 

\begin{proposition}{1.1}
For $N$ independent log-normally distributed patterns $\{\mathbf{x}^1, ...,\mathbf{x}^N \}$ with $ \ln \mathbf{x}^k_i \sim \mathcal{N}(\mathbf{\mu}^k_i, (\sigma^k_i)^2) $, the probability distribution of the synaptic connection $w_{ij}:=\sum_{k=1}^N\mathbf{x}^k_i\mathbf{x}^k_j$ in the Hebbian network can be approximated using a new log-normal distribution denoted by: 
\begin{equation}
\ln w_{ij} \overset{\text{approx.}}{\sim}\mathcal{N}(\mathbf{\mu}_{w_{ij}}, (\sigma_{w_{ij}})^2) 
\end{equation}
i.e., 

\begin{equation}
    p(w_{ij}) \approx \frac{1}{w_{ij}\sigma_{w_{ij}} \sqrt{2\pi}} \exp\left(
    -\frac{(\ln w_{ij} - \mu_{w_{ij}})^2}{2\sigma_{w_{ij}}^2}
    \right)
\label{eq: pwij}
\end{equation}
with the parameters $\mu_{w_{ij}}$ and $ \sigma_{w_{ij}}^2$ being:
\begin{equation}
     \mu_{w_{ij}} = \ln M_1 - \frac{1}{2}\ln(1 + \frac{M_2}{M_1^{\ 2}}), \quad \sigma_{w_{ij}}^2 = \ln(1 + \frac{M_2}{M_1^{\ 2}})
\label{eq: mu, sig single w}
\end{equation}
and with $M_1$ and $M_2$ defined as:
\begin{align}
    &M_1 := \sum_{k=1}^N \exp\left(\mathbf{\mu}^k_i + \mathbf{\mu}^k_j + 
    \frac{1}{2}\left[(\sigma^k_{i})^2 + (\sigma^k_{j})^2 + 2\sigma^k_{ij}\right] \right)
    \label{eq: M1 single w}\\
    &M_2 := \sum_{k=1}^N \left[ \exp\left((\sigma^k_{i})^2 + (\sigma^k_{j})^2 + 2\sigma^k_{ij}\right) - 1 \right]  \exp\left(2 (\mathbf{\mu}^k_i + \mathbf{\mu}^k_j) +
    (\sigma^k_{i})^2 + (\sigma^k_{j})^2 + 2\sigma^k_{ij} \right)
\label{eq: M2 single w}
\end{align}
\end{proposition}

Next, we derive the expression for the conditional distribution $ p(w_{ij} | \mathbf{x}^l) $. Given the independence of the $ N $ terms defining a given synaptic weight, i.e., $ \mathbf{x}^k_i\mathbf{x}^k_j \perp \mathbf{x}^l_i\mathbf{x}^l_j \ \forall k \neq l $, the distributions $ p(w_{ij} | \mathbf{x}^l) $ and $ p\left(\sum_{k \neq l}^N \mathbf{x}^k_i \mathbf{x}^k_j\right) $ are identical in the sense that the latter is a shifted version of the former by a constant $ -\mathbf{x}^l_i \mathbf{x}^l_j $. For convenience, we denote $w_{ij/l}:=\sum_{k \neq l}^N \mathbf{x}^k_i \mathbf{x}^k_j$. Consequently, we can derive the probability distribution for $ w_{ij/l} $ instead, and do so using the same derivation steps we used in Proposition 1.1 for $ w_{ij}$. We denote this distribution by:
\begin{equation}
    \ln w_{ij/l} := \ln\left(\sum_{k \neq l}^N \mathbf{x}^k_i \mathbf{x}^k_j\right) \overset{\text{approx.}}{\sim}\mathcal{N}\left(\mu_{w_{ij/l}}, (\sigma_{w_{ij/l}})^2\right),
\label{eq: wij given l}
\end{equation}
where the parameters $ \mu_{w_{ij/l}} $ and $ (\sigma_{w_{ij/l}})^2 $ have the same form as in Equation \ref{eq: mu, sig single w}, with the $ l^\mathrm{th} $ term removed from the summation in Equations \ref{eq: M1 single w} and \ref{eq: M2 single w}. For detailed expressions, see section \ref{sec: proofs}.

With both $ p(w_{ij}) $ and $ p(w_{ij} | \mathbf{x}^l) $ known, we can now obtain the expression for the mutual information.

\begin{proposition}{1.2}
The mutual information between a synaptic connection and a data pattern is approximated by:
\begin{equation}
    MI(w_{ij}; \mathbf{x}^l) = \left(\mu_{w_{ij}} - \mu_{w_{ij/l}}\right) + \left(\ln \sigma_{w_{ij}} - \ln \sigma_{w_{ij/l}}\right),
\label{eq: MI single conn}
\end{equation}
with the units in nats. 
\end{proposition}

We note that the mutual information between a synaptic connection $w_{ij}$ and a data pattern $\mathbf{x}^l$ (eqn. \ref{eq: MI single conn}), depends on the specific connection weight $w_{ij}$
(eqn. \ref{eq: pwij}), the specific pattern $\mathbf{x}^l$, and the distribution properties of all data patterns, specifically their means and covariance matrices (eqns. \ref{eq: mu, sig single w}, \ref{eq: M1 single w},  \ref{eq: M2 single w}).
 
\subsection{Information encoded by a pair of connections}
Since neural firing activities within a network can be correlated, the corresponding connection weights are likely to exhibit statistical dependencies rather than being independent of one another. To study the collective coding by these connections, we focus first on the mutual information between data and pairs of connections: $ MI(\mathbf{w}; \mathbf{x}^l) $, where $ \mathbf{w} = (w_{ij}, w_{mn})^T $. This analysis requires knowledge of the distributions $ p(\mathbf{w}) $ and $ p(\mathbf{w} | \mathbf{x}^l) $.

From the derivation in the single connection case, we obtain the marginals for each weight, which are given by:
\begin{equation}
\ln w_{ij} \overset{\text{approx.}}{\sim} \mathcal{N}\left(\mu_{w_{ij}}, (\sigma_{w_{ij}})^2\right), \quad \ln w_{mn} \overset{\text{approx.}}{\sim} \mathcal{N}\left(\mu_{w_{mn}}, (\sigma_{w_{mn}})^2\right),
\end{equation}
This result indicates that the individual components of the vector \( \ln \mathbf{w} = (\ln w_{ij}, \ln w_{mn})^T \) are Gaussian. While this does not necessarily imply that the joint distribution is multivariate Gaussian, we follow a common approximation for tractability by modeling the joint distribution as bivariate Gaussian here. Specifically, we express the joint distribution of \( \ln \mathbf{w} \) as:
\begin{equation}
\renewcommand{\arraystretch}{0.6} 
\setlength{\arraycolsep}{3pt} 
\ln \mathbf{w} = 
\begin{pmatrix}
\ln w_{ij} \\
\ln w_{mn}
\end{pmatrix} \overset{\text{approx.}}{\sim} \mathcal{N} (
\mu_{\mathbf{w}} = 
\begin{pmatrix}
\mu_{w_{ij}} \\
\mu_{w_{mn}}
\end{pmatrix}, 
\Sigma_\mathbf{w} =
\begin{pmatrix}
\sigma_{w_{ij}}^2 & \sigma_{w_{ij, mn}} \\
\sigma_{w_{mn,ij}} & \sigma_{w_{mn}}^2
\end{pmatrix}
).
\end{equation}
The only remaining unknown in this equation is $ \sigma_{w_{ij, mn}} $, which represents the covariance between $ \ln w_{ij} $ and $ \ln w_{mn} $. The expression for this covariance is presented in the following proposition:

\begin{proposition}{2.1}
Under the approximation that a pair of synaptic connections $(w_{ij}, w_{mn})^T $ $:= (\sum_k^N \mathbf{x}^k_i \mathbf{x}^k_j, \sum_k^N \mathbf{x}^k_m \mathbf{x}^k_n)^T$ follows a 2-dimensional log-normal distribution, the covariance between $\ln w_{ij}$ and $\ln w_{mn}$ is:
\begin{align}
&\sigma_{w_{ij,mn}} = \ln ( 1 \; + \nonumber \\ 
&\frac{\sum_{k=1}^N \exp\left(
\sum_{\phi}^{\{i,j,m,n\}} \mu^k_{\phi} 
+ \frac{1}{2}\sum_{\phi}^{\{i,j,m,n\}} (\sigma^k_{\phi})^2
+ \sum_{\phi}^{\{ij,mn\}} \sigma^k_{\phi}
\right)\left[\exp\left( 
\sum_{\phi}^{\{i,j\}\times\{m,n\}} \sigma^k_{\phi}
\right) - 1\right]}
{\sum_{k,l=1}^N \exp\left( 
\sum_{\phi}^{\{i,j\}} \mu^k_{\phi} 
+ \sum_{\phi}^{\{m,n\}} \mu^l_{\phi} 
+ \frac{1}{2}\sum_{\phi}^{\{i,j\}} (\sigma^k_{\phi})^2
+ \frac{1}{2}\sum_{\phi}^{\{m,n\}} (\sigma^l_{\phi})^2
+ \sigma_{ij}^k + \sigma_{mn}^l
\right)}
)
\label{eq: cov lnwij lnwmn}
\end{align}
with the summation notation meaning:
\begin{align}
\sum_{\phi}^{\{ij,mn\}} \sigma^k_{\phi} = \sigma^k_{ij} +  \sigma^k_{mn} \; ,
\sum_{\phi}^{\{i,j\}\times\{m,n\}} \sigma^k_{\phi} = \sigma^k_{im} + \sigma^k_{in} + \sigma^k_{jm} + \sigma^k_{jn}
\end{align}
\end{proposition}

With $ p(\mathbf{w}) $ thus specified, we next turn to $ p(\mathbf{w} | \mathbf{x}^l) $. As was the case 
with single connections, here, $ p(\mathbf{w} | \mathbf{x}^l) $ and $ p\left(\sum_{k\neq l}^N \mathbf{x}^k_i\mathbf{x}^k_j, \sum_{k\neq l}^N \mathbf{x}^k_m\mathbf{x}^k_n\right) $ are identical distributions up to a location shift. For convenience, we denote $\mathbf{w}_{/l}:=(\sum_{k\neq l}^N \mathbf{x}^k_i\mathbf{x}^k_j, \sum_{k\neq l}^N \mathbf{x}^k_m\mathbf{x}^k_n)^T$. Applying the same steps we used to derive $ p(\mathbf{w}) $ above, we get:
\begin{equation}
\renewcommand{\arraystretch}{0.6} 
\setlength{\arraycolsep}{3pt} 
\ln \mathbf{w}_{/l} := 
\begin{pmatrix}
\ln w_{ij/l} \\
\ln w_{mn/l}
\end{pmatrix} \overset{\text{approx.}}{\sim} \mathcal{N} (
\mu_{\mathbf{w}_{/l}} = 
\begin{pmatrix}
\mu_{w_{ij/l}} \\
\mu_{w_{mn/l}}
\end{pmatrix}, 
\Sigma_{\mathbf{w}_{/l}} =
\begin{pmatrix}
\sigma_{w_{ij/l}}^2 & \sigma_{w_{ij, mn/l}} \\
\sigma_{w_{mn,ij/l}} & \sigma_{w_{mn/l}}^2
\end{pmatrix}
),
\end{equation}
where the parameters with $ /l $ are equal to their original corresponding parameters, with the terms related to $ \mathbf{x}^l $ removed from the summation. For a more detailed expression for $\mu_{\mathbf{w}_{/l}}$ and $ \Sigma_{\mathbf{w}_{/l}}$, please refer to the Methods section \ref{sec: proofs}.

Given $ p(\mathbf{w}) $ and $ p(\mathbf{w} | \mathbf{x}^l) $, we can then calculate the mutual information encoded by any pair of synaptic connections about a data pattern, as stated in the following proposition:

\begin{proposition}{2.2}
The mutual information between a pair of connections $ \mathbf{w} = (w_{ij}, w_{mn})^T $ and a data pattern $ \mathbf{x}^l $ is given by the analytical approximation:
\begin{equation}
MI(\mathbf{w}; \mathbf{x}^l) = (\mu_{w_{ij}} + \mu_{w_{mn}} - \mu_{w_{ij/l}} - \mu_{w_{mn/l}}) + \frac{1}{2}\ln|\Sigma_\mathbf{w}| - \frac{1}{2}\ln|\Sigma_{\mathbf{w}_{/l}}|,
\end{equation}
where $ |\Sigma_\mathbf{w}| $ denotes the determinant of the matrix $ \Sigma_\mathbf{w} $. The information is expressed in nats.
\end{proposition}

\subsection{Information in an ensemble of multiple synaptic connections}
Next, we extend the analysis to information encoded jointly in an ensemble of $ n $ synaptic connections, denoted as $ \mathbf{w} = (w_{ij(1)}, w_{ij(2)},\dots, w_{ij(n)})^T $. Here, the set $\{ij(1), ij(2),$ $\dots, ij(n)\}$ represents the indices for the $ n $ synaptic weights. For example, $ w_{ij(k)} $ refers to the connection between neuron $ i(k) $ and neuron $ j(k) $. The mutual information encoded by any $ n $ connections about the data, $ MI(\mathbf{w}; \mathbf{x}^l) $, can be derived using a similar approach as before. Since all marginals for $ \ln \mathbf{w} $ are normally distributed, we approximate the joint distribution as an $ n $-dimensional log-normal distribution, leading to:
\begin{equation}
\ln \mathbf{w} =
\renewcommand{\arraystretch}{0.6} 
\setlength{\arraycolsep}{2pt} 
\begin{pmatrix}
\ln w_{ij(1)} \\
\vdots\\
\ln w_{ij(n)}
\end{pmatrix} \overset{\text{approx.}}{\sim} \mathcal{N} (
\mu_\mathbf{w} = 
\begin{pmatrix}
\mu_{w_{ij(1)}} \\
\vdots \\
\mu_{w_{ij(n)}}
\end{pmatrix},
\Sigma_\mathbf{w} =
\begin{pmatrix}
\sigma_{w_{ij(1)}}^2 & \sigma_{w_{ij(1),ij(2)}} & \hdots &\sigma_{w_{ij(1), ij(n)}} \\
\sigma_{w_{ij(2),ij(1)}} & \sigma_{w_{ij(2)}}^2 & \hdots &\sigma_{w_{ij(2), ij(n)}} \\
\vdots & \vdots & \ddots & \vdots \\
\sigma_{w_{ij(n),ij(1)}} & \sigma_{w_{ij(n),ij(2)}} & \hdots & \sigma_{w_{ij(n)}}^2
\end{pmatrix}
)
\end{equation}
This expression represents $ p(\mathbf{w}) $ as an $ n $-dimensional log-normal distribution. The parameters $ \mu_{\mathbf{w}} $ and the diagonal elements of $ \Sigma_{\mathbf{w}} $ follow the form given in Equation \ref{eq: mu, sig single w}, while the off-diagonal elements of $ \Sigma_{\mathbf{w}} $ follow Equation \ref{eq: cov lnwij lnwmn}, specifically:
\begin{align}
&\sigma_{w_{ij(r),ij(s)}} = \ln ( 1 \; + \nonumber \\ 
&\frac{\sum_{k=1}^N \exp\left(
\sum_{m}^{\{r,s\}} 
\left[\sum_{\phi}^{\{i(m),j(m)\}}
\mu^k_{\phi} 
+ \frac{1}{2} (\sigma^k_{\phi})^2\right]
+ \sigma^k_{ij(m)}
\right)\left[\exp\left( 
\sum_{\phi}^{\{i(r),j(r)\}\times\{i(s),j(s)\}} \sigma^k_{\phi}
\right) - 1\right]}
{\sum_{k,l=1}^N \exp\left( 
\left[\sum_{\phi}^{\{i(r),j(r)\}} \mu^k_{\phi} 
+ \frac{1}{2}(\sigma^k_{\phi})^2\right]
+ \;   \left[\sum_{\phi}^{\{i(s),j(s)\}} \mu^l_{\phi} 
+ \frac{1}{2} (\sigma^l_{\phi})^2\right]
+ \sigma_{ij(r)}^k + \sigma_{ij(s)}^l
\right)}
)
\end{align}

To obtain $ p(\mathbf{w} | \mathbf{x}^l) $, as in the single and paired connection scenarios, we focus on its shifted distribution $ p(\mathbf{w}_{/l}) $ with $\mathbf{w} = (w_{ij(1)/l}, w_{ij(2)/l}, \dots, w_{ij(n)/l})^T$ , which is also treated as log-normally distributed. The parameters with $ /l $ are simply the original parameters with the summation terms related to $ \mathbf{x}^l $ excluded:
\begin{equation}
\ln \mathbf{w}_{/l} =
\renewcommand{\arraystretch}{0.6} 
\setlength{\arraycolsep}{3pt} 
\begin{pmatrix}
\ln w_{ij(1)/l} \\
\vdots\\
\ln w_{ij(n)/l}
\end{pmatrix} \overset{\text{approx.}}{\sim} \mathcal{N} (
\mu_{\mathbf{w}_{/l}} = 
\begin{pmatrix}
\mu_{w_{ij(1)/l}} \\
\vdots \\
\mu_{w_{ij(n)/l}}
\end{pmatrix}, 
\Sigma_{\mathbf{w}_{/l}} =
\begin{pmatrix}
\sigma_{w_{ij(1)/l}}^2  & \hdots &\sigma_{w_{ij(1), ij(n)/l}} \\
\vdots  & \ddots & \vdots \\
\sigma_{w_{ij(n),ij(1)/l}} & \hdots & \sigma_{w_{ij(n)/l}}^2
\end{pmatrix}
)
\end{equation}
Please refer to section \ref{sec: proofs} for detailed expressions for $\mu_{\mathbf{w}_{/l}}$ and $\Sigma_{\mathbf{w}_{/l}}$.

Finally, with $ p(\mathbf{w}) $ and $ p(\mathbf{w} | \mathbf{x}^l) $, we can derive the expression for the information jointly encoded in multiple synaptic connections about a pattern:

\begin{proposition}{3}
The mutual information between a collect of $n$ synaptic connections $\mathbf{w}=(w_{ij(1)}, w_{ij(2)}, ..., w_{ij(n)})^T$ and a data pattern $\mathbf{x}^l$ is given by the analytical approximation: 
\begin{equation}
     MI(\mathbf{w};\mathbf{x}^l) = \sum_{k=1}^n \left(\mu_{w_{ij(k)}}  - \mu_{w_{ij(k)/l}}\right) + \frac{1}{2}\ln|\Sigma_\mathbf{w}| - \frac{1}{2}\ln|\Sigma_{\mathbf{w}_{/l}}| \label{eq: MI ensemble}
\end{equation}
\end{proposition}

Together, these derivations provide an explicit expression for the information encoded in a specified synaptic ensemble about a particular data pattern, 
$MI(\mathbf{w};\mathbf{x}^l)$. Notably, it 
is fully determined by the parameters of the distributions 
(means and covariances) of the relevant components of all $N$ data patterns.  Formally, this dependence can be summarized by the set $\mathcal{S} = \{\mu^k_i, \sigma^k_{ij} \mid k \in \{1, \ldots, N\}, \forall (i, j) \text{ such that }$ $w_{ij} \text{ is a component of } \mathbf{w}\}$. Consequently, both the size of the ensemble, $n$, and the particular choice of connections included in the ensemble, influence $MI$, as they determine the elements included in the set $\mathcal{S}$.

Importantly, $MI(\mathbf{w};\mathbf{x}^l)$ depends only on the parameters associated with the neurons participating in the ensemble; neurons in the network that lie outside the ensemble have no effect. This finding implies that as long as the local statistics, $\mathcal{S}$, of the specified neurons remain fixed, the value of $MI(\mathbf{w};\mathbf{x}^l)$ will remain unchanged, regardless of the larger Hebbian connectivity in which the ensemble is embedded. In this sense, our result is context-invariant: it depends solely on the properties of the synaptic ensemble in question and is independent of the total size, $d$, of the network containing the ensemble.

\section{Results}
After obtaining the analytical expressions for the information encoded in synaptic connections, we performed a series of exploratory analyses by varying the parameter set $\mathcal{S}$ to examine its effect on $MI$.  

We conducted three sets of simulations. The first set varied parameter values in constrained cases to reveal the dependency of $MI$ on continuously changing parameters. The second and third sets relaxed these constraints progressively and sampled parameter combinations randomly; regression analysis was used to reveal the dependency of $MI$ on a range of parameter combinations. Below, we describe the key differences in the constraints used for each simulation set to facilitate understanding of the subsequent results. For further details regarding the simulation setups, please see the Methods section.

\noindent \textbf{First Simulation Set:} The covariance matrix $\Sigma^k$ was held constant across different data patterns, and was constrained to be exchangeable: all diagonal elements were set to $\tilde{\sigma}^2$, and off-diagonal elements to $\tilde{\rho} \tilde{\sigma}^2$. Results of this analysis are shown in Figures 2A, 2B, 3A, and 3B.

\noindent \textbf{Second Simulation Set:} 
Again, $\Sigma^k$ was held constant across data patterns. However, to explore more general covariance structures, a stochastic component was added to the exchangeable component (making them no longer exchangeable); it was set to be proportional to $\mathbf{A}^T\mathbf{A}$, where $\mathbf{A}$ is a stochastic matrix. Results are shown in Figures 2C and 3C.

\noindent \textbf{Third Simulation Set:} 
Here, the covariance matrix $\Sigma^k$ varied between patterns. To avoid biases in matrix generation, each $\Sigma^k$ was constructed by generating random eigenvalues and random orthogonal matrices. Results of the corresponding analyses are shown in Figures 2D, 3D, and 4.

We note that we use nats as the unit of mutual information in both the theoretical section and simulations for consistency.

\subsection{Relationship between the information encoded by synapses and the number of data patterns}

{
\begin{center}
\includegraphics[width=0.9\textwidth]{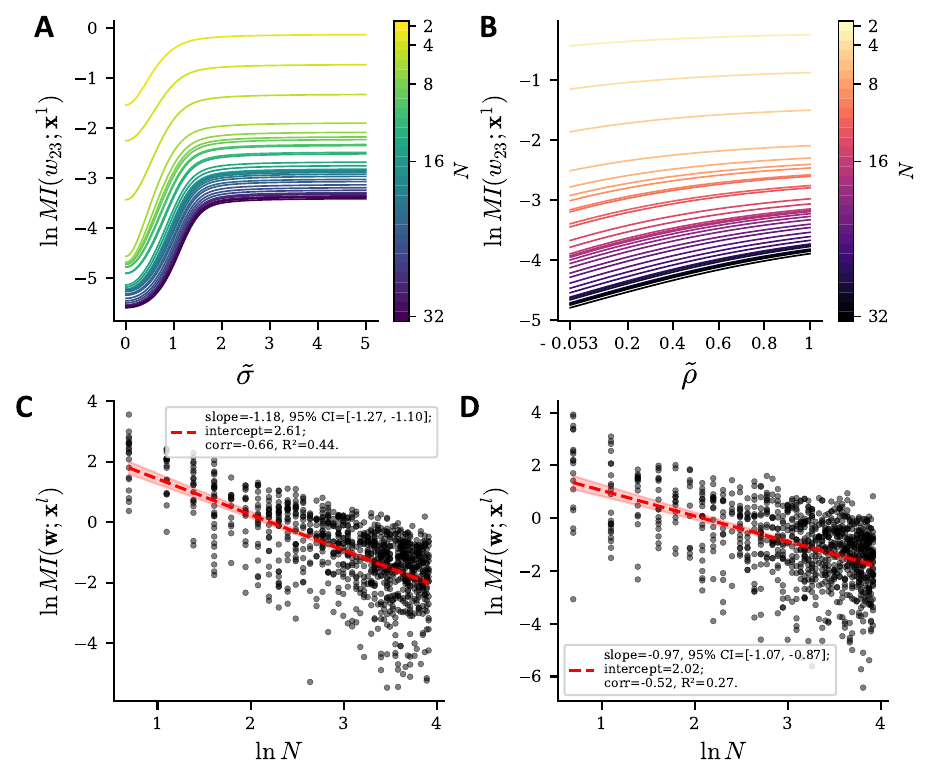}
\end{center}
\captionof{figure}{Encoded information as a function of the number of data patterns $N$.
(A,B) Results from the first simulation set, and for a \textit{single} (randomly chosen) synaptic connection. (A) Information encoded about a specific example pattern ($ \mathbf{x}^1 $) by a single example weight ($ w_{23} $), plotted as a function of $ \tilde{\sigma} $ (x-axis) and the number of data patterns $ N $ (different colors). $ \tilde{\rho} $ is fixed at 0, indicating that the dimensions of the data patterns are independent.  
(B) Information encoded about $ \mathbf{x}^1 $ by $ w_{23} $, plotted as a function of $ \tilde{\rho} $ (x-axis) and $ N $ (colors); $ \tilde{\sigma} $ is fixed at 1. 
In both (A) and (B), the information encoded decreases as $ N $ increases, indicated by downward shifts in the curves for higher $ N $.
(C, D) Results from the second and third simulation sets, respectively, and for synaptic ensembles $\mathbf{w}$. The second simulation set corresponds to data patterns with the same, but non-exchangeable covariance matrix, and the third simulation set to data patterns with different, random covariance matrices. Each black dot corresponds to a random choice of configuration, including the number ($n$) and identity of synaptic connections in the ensemble, the distribution of data patterns, and the specific data pattern $ \mathbf{x}^l $. $n$ ranges from 1 to 20. The red dashed line indicates the fitted linear regression, with the red shaded area representing the $95\%$ confidence interval for the predicted values. The decreasing trend of $MI$ with increasing $ N $ also persists for synaptic \textit{ensembles} $ \mathbf{w} $ with randomly sampled configurations.\\}
\label{Fig:MI and pattern number}
}

We investigated the relationship between the information encoded by synapses and the number of patterns $N$ in the dataset. 

First, to develop intuition for this relationship, we began by examining how mutual information changes with $N$: (i) for \textit{single} synaptic connections, here, randomly chosen to be $w_{23}$, and (ii) under the simplest data covariance conditions, here, using the first simulation set. We plotted $MI(w_{23};\mathbf{x}^1)$ as a function of $\tilde{\sigma}$ (Fig. \ref{Fig:MI and pattern number}A), and as a function of $\tilde{\rho}$ (Fig. \ref{Fig:MI and pattern number}B), and in each case, obtained curves for varying numbers of data patterns ($N$, ranging from 2 to 32). 

We found that the information encoded in a single connection increases with both $\tilde{\sigma}$ and $\tilde{\rho}$ (Fig. \ref{Fig:MI and pattern number}AB), appearing to reach an asymptotic limit for large $\tilde{\sigma}$. These findings are consistent with the explanation that because $\tilde{\sigma}$ represents the spread or width of the distribution, a wider distribution, typically corresponding to higher entropy in the data patterns, provides more content for the network to encode. Meanwhile, $\tilde{\rho}$ reflects the internal structure of the data: data with more structure (lower $\tilde{\rho}$) tends to be easier (lower intrinsic dimensionality) for the network to store, assuming the same level of entropy. 

Notably, we found that as the number of data patterns increases from 2 to 32, the information encoded by a single connection $MI(w_{23};\mathbf{x}^1)$ decreases independently of the specific $\tilde{\sigma}$ and $\tilde{\rho}$ values  (Fig. \ref{Fig:MI and pattern number}AB; downward shift of curves from light green/light red to dark purple; log scale). 
This indicates that the information stored in each individual connection about a single data pattern decreases as the total number of patterns to be stored in the neural network increases, rather than remaining constant.

Next, with this intuition in hand, we relaxed the above constraints to consider the more general scenario of ensemble coding. Specifically, we examined how  $MI$ changes with $N$: (i) for synaptic \textit{ensembles} of random sizes, here, between $n=1$ to $n=20$, and (ii) under more general covariance structures, i.e., using the second and third simulation sets in which the covariance matrices are not restricted to be exchangeable.

In Fig. \ref{Fig:MI and pattern number}C, where the (non-exchangeable) covariance matrices are identical across patterns, we observed a significant negative correlation between $MI(\mathbf{w};\mathbf{x}^l)$ and $N$ (slope=-1.18, 95\% CI=[-1.27, -1.10]; $R^2=0.44$).  
Similarly, in Fig. \ref{Fig:MI and pattern number}D, where the covariance matrices differed among patterns, we observed a significant negative correlation (slope=-0.97, 95\% CI=[-1.07, -0.87]; $R^2=0.27$). 
Furthermore, by examining the slope of the linear relationship between $\ln MI(\mathbf{w}; \mathbf{x}^l)$ and $\ln N$, we find that $MI(\mathbf{w}; \mathbf{x}^l)$ follows an approximate $N^{-1.2}$ scaling in Fig. \ref{Fig:MI and pattern number}C and $N^{-1.0}$ in Fig. \ref{Fig:MI and pattern number}D (with accuracy kept to the first decimal place). 
Thus, our results revealed that the information encoded in synaptic ensembles, in general, decreases with the number of data patterns. 

\subsection{Relation between information encoded by synapses and the number of synaptic connections}

Next, we investigated how mutual information changes in relation to the number of synaptic connections, rather than the number of data patterns.

Results from the first simulation set provided intuition for how the information encoded about a data pattern by a synaptic ensemble, i.e.,  $MI(\mathbf{w};\mathbf{x}^1 )$, changes as a function of $\tilde{\sigma}$ (Fig. \ref{Fig:MI & ensemble size}A), and as a function of $\tilde{\rho}$ (Fig. \ref{Fig:MI & ensemble size}B).
Curves for varying numbers of synaptic connections ($n$, ranging from 1 to 16) are overlaid for comparison. 

{
\begin{center}
\includegraphics[width=0.9\textwidth]{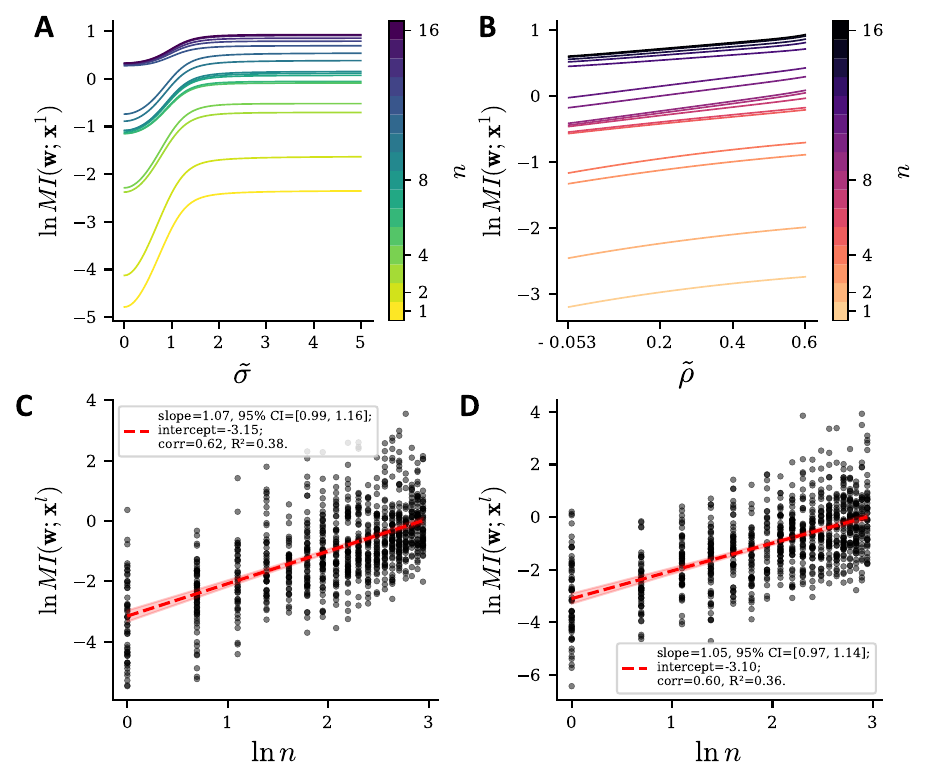}
\end{center}
\captionof{figure}{Encoded information as a function of the number of synaptic connections in an ensemble. 
(A, B) Results from the first simulation set. A randomly chosen ensemble ($ \mathbf{w} $) and a randomly selected example pattern ($ \mathbf{x}^1 $) are used to illustrate how the mutual information depends on $ \tilde{\sigma} $, $ \tilde{\rho} $, and the number of synaptic connections $ n $.  
(A) Information encoded about $ \mathbf{x}^1 $ by the example ensemble, plotted as a function of $ \tilde{\sigma} $ (x-axis) for varying $ n $ (colors). Here, $ \tilde{\rho} = 0 $ and $ N = 10 $.   (B) Information encoded about $ \mathbf{x}^1 $ by the example ensemble, plotted as a function of $ \tilde{\rho} $ (x-axis) for different values of $ n $ (colors). Here, $ \tilde{\sigma} = 1 $ and $ N = 10 $. In both (A) and (B), as $ n $ increases, the curves shift upward, indicating that larger ensembles encode more information.  
(C, D) Results from the second and third simulation sets, respectively. Other conventions as in Figure \ref{Fig:MI and pattern number}CD. The increasing trend of $MI$ as a function of $n$ persists when both $ \mathbf{w} $ and all parameter configurations are  sampled randomly. \\}
\label{Fig:MI & ensemble size}
}

Similar to the simple case of a single synaptic connection (Fig. \ref{Fig:MI and pattern number}AB), the increasing relationship of $MI(\mathbf{w}; \mathbf{x}^1)$ with $\tilde{\sigma}$ and $\tilde{\rho}$ holds true for an ensemble of synaptic connections as well (Fig. \ref{Fig:MI & ensemble size}AB).

Notably, we found that as the number of synaptic connections increases from 1 to 16, the information in the ensemble $MI(\mathbf{w};\mathbf{x}^1 )$ also increases (Fig. \ref{Fig:MI & ensemble size}AB; upward shift of curves from light green/light red to dark purple; log scale). This result is consistent with the notion that adding more variables should allow for more of the information about the patterns to be encoded.

This increasing relationship was evident also in the second  (Fig. \ref{Fig:MI & ensemble size}C) and third (Fig. \ref{Fig:MI & ensemble size}D) simulation sets, involving progressively more general covariance structures. In Fig. \ref{Fig:MI & ensemble size}C, where the covariance matrices are non-exchangeable but identical across patterns, we observed a significant positive correlation between $MI(\mathbf{w};\mathbf{x}^l)$ and $n$ (slope=1.07, 95\% CI=[0.99, 1.16]; $R^2=0.38$). Similarly, in Fig. \ref{Fig:MI & ensemble size}D, where the covariance matrices differed among patterns, we observed a significant positive correlation between $MI(\mathbf{w};\mathbf{x}^l)$ and $n$ (slope=1.05, 95\% CI=[0.97, 1.14]; $R^2=0.36$).
    
Interestingly, these results allow us to estimate the relationship between the number of data patterns, $ P $, stored in the entire network and the size of the network (or number of neurons), $ d $. 
First, linear regression between $MI$ and the number $n$ of synaptic connections in the ensemble reveal a power-law relationship of $ MI(\mathbf{w}; \mathbf{x}^l) \sim n^{1.1} $ (with accuracy kept to the first decimal place), which is nearly identical across simulation sets (Figs. \ref{Fig:MI & ensemble size}CD). In other words, the information capacity of synaptic ensembles scales approximately with the 1.1 power of their size. Second, since our modeled network is an undirected graph without self-connections, the size of the full synaptic ensemble and the size of the network are related quadratically, i.e., $ n = d(d-1)/2 \sim d^2 $. Therefore, the information capacity of synaptic ensembles scales as $ MI \sim d^{2.2} $. Third, in our autoassociation setup, all neurons function both as inputs and outputs for memorization, with no hidden neurons involved. Thus, the network size equals the dimensionality of each data pattern. If we define successful memorization as the ability to accurately reconstruct each component of a pattern within the same resolution, then each component requires an equal amount of information for storage. Accordingly, the information required to store one pattern, $H_{\mathrm{pattern}}$, is proportional to the number of components, i.e., $ H_{\mathrm{pattern}} \sim d $. 
Putting these together, we obtain the number of patterns that can be stored in the network by dividing the total information stored in the network by the information needed to store a single pattern: 
\begin{equation*}
    P \approx \frac{MI}{H_{\mathrm{pattern}}} \sim d^{1.2},
\end{equation*}
This yields that the number of patterns that can be stored in the connectivity of a dense Hebbian network is proportional to the 1.2 power of the network size.

\subsection{Information contributed per connection increases with the synaptic ensemble size}

{
\begin{center}
\includegraphics[width=0.9\textwidth]{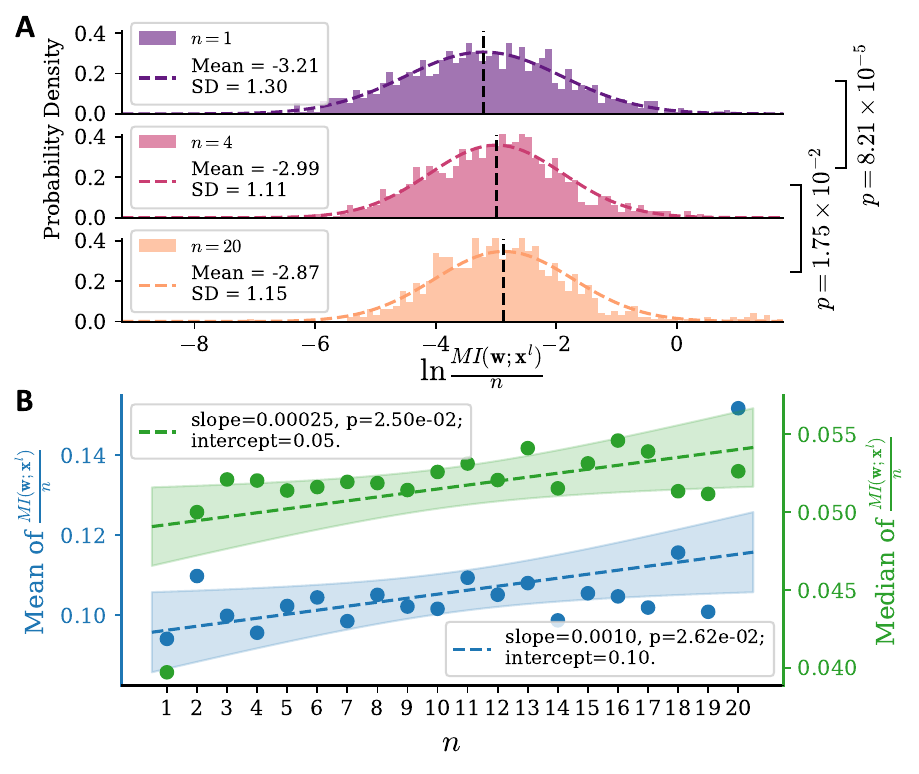}
\end{center}
\captionof{figure}{Information contributed per synaptic connection as a function of ensemble size. (A) Histograms showing the distribution of $ \ln(MI/n) $ for $ n = 1, 4, 20 $.
(B) Relationship between the mean (blue) and median (green) of $MI/n$ for different values of $n$. Dashed lines indicate the linear regression results for these mean and median values, while shaded areas indicate the $95\%$ confidence intervals for the predicted values.\\
}
\label{Fig:Synergy}
}

In the previous sections, we analyzed the information encoded by a single synaptic connection, $ MI(w; \mathbf{x}^l) $, and by an ensemble of synaptic connections, $ MI(\mathbf{w}, \mathbf{x}^l) $. However, an intriguing question that arises is: how does the contribution of a single connection to the information encoded by the entire synaptic ensemble change with the size of the  ensemble? To address this question, we approximated the contribution of a each connection to the information encoded by the ensemble as $ MI(\mathbf{w}, \mathbf{x}^l) / n $, where $ n $ is the size of the ensemble. This analysis sheds light on how individual connections behave when functioning as part of a collective.

Using data from simulation set three, which reflects the most general configuration, we computed the relationship between $ \ln (MI(\mathbf{w}, \mathbf{x}^l) / n) $ and $ n $. We first plotted the histograms of $ \ln (MI(\mathbf{w}, \mathbf{x}^l) / n) $ for $n=$ 1, 4, and 20, and found that they followed Gaussian-like distributions (Fig. \ref{Fig:Synergy}A). Interestingly, this Gaussian-like shape suggests that the information contribution per synapse may also follow a log-normal distribution. Furthermore, these distributions shifted toward higher values
as $ n $ increases; the shifts were significant between $ n = 1 $ and $ n = 4 $ ($p = 8.21\times 10^{-5}$, two-sided t-test), and between $ n = 4 $ and $ n = 20 $ ($p = 1.75\times 10^{-2}$, two-sided t-test) indicating notable changes in the distribution. 

To explore this relationship further, we examined the values of $MI(\mathbf{w}, \mathbf{x}^l) / n$ as a function of $n$ and found significant \textit{positive} correlation (slope=$9.6 \times 10^{-4}$, 95\% CI=[$4.3\times 10^{-4}, 1.5\times 10^{-3}$]). Furthermore, we also examined the median and median values of $MI(\mathbf{w}, \mathbf{x}^l) / n$. There is a significant upward trend for both the mean (Fig. \ref{Fig:Synergy}B; blue line, $p = 2.62\times 10^{-2}$) and the median (Fig. \ref{Fig:Synergy}B; green line, $p = 2.50\times 10^{-2}$) as $n$ increases. These findings reveal that the information contributed per synaptic connection increases with ensemble size.

(B) Data points showing the relationship between $MI/n$ and $n$. The red dashed line represents the linear regression fit, and the red shaded area indicates the $95\%$ confidence interval for the predicted values. Note that $MI/n$ is shown on a logarithmic scale.

\section{Discussion}
In this study, we adopted an under-explored perspective on neural coding, focusing on the information encoded in synaptic weights rather than in neuronal firing activities. 
Our objective was to derive an analytical expression for the information embedded in ensembles of synaptic connections. 
As a first step in addressing this gap, here, we used a network with Hebbian connectivity to derive a closed-form expression for the mutual information between log-normally distributed data patterns and ensembles of synaptic weights.
Although seemingly complex, the resulting formulation represents a straightforward analytical approximation involving just the statistical parameters of the distributions of the data patterns. Using this closed-form expression, we conducted three sets of simulations that revealed interesting insights into distributed coding, network capacity, and information synergy, which are discussed in detail in the following sections. Overall, we propose that our work provides a framework for analyzing neural information through the lens of synaptic connections, offering a complementary perspective to the traditional focus on firing activities. 

\subsection{Distributed coding in synaptic weights}
We demonstrated, here, that the mutual information about a data pattern, whether stored in a single synaptic weight, $ MI(w_{ij}; \mathbf{x}^1) $, or an ensemble of synaptic weights, $ MI(\mathbf{w}; \mathbf{x}^l) $, decreases as the number of stored patterns, $ N $, increases. This suggests that introducing new patterns for storage reduces the capacity of synaptic connections to retain information about previously stored patterns. In other words, each synaptic connection allocates part of its storage capacity to every new pattern, resulting in a scenario in which individual connections simultaneously encode partial information about all stored patterns.  
In contrast, one could imagine an alternative scheme, analogous to the “grandmother cell” hypothesis in neural activity coding, in which information storage relies on “grandmother connections”: certain synapses specialize in encoding specific patterns, while others specialize in different patterns. In this scenario, introducing new patterns would only use capacity in previously unused connections, leaving the information stored in other connections unaffected. 
Our findings, however, align better with the well-known paradigm of distributed coding in the brain, in which diverse types of information—such as memory \citep{rissman2012distributed}, sensory inputs \citep{stecker2003distributed}, or motor actions \citep{steinmetz2019distributed}—are encoded collectively across groups of neurons, and further extend this principle to the perspective of synaptic connections within neural networks.  

\subsection{Capacity analysis for networks with Hebbian connectivity}

Our work quantifies the capacity of Hebbian networks based on the information stored within the synaptic ensemble. Since Hebbian connectivity is fundamental to the Hopfield network, we use the latter as a reference for comparison. The capacity of the Hopfield network has been extensively explored, with well-known results such as $P \sim 0.14d$ \citep{amit1985storing} or $P \leq d$ \citep{abu1985information}. Our analysis shows that a distinct storage law applies to Hebbian networks considered broadly. By assuming that each component or ``digit" in a pattern requires the same amount of information, we found that the number of stored patterns ($P$) follows a marginally superlinear relationship with the network size ($d$), specifically $P \sim d^{1.2}$.

This difference can be attributed to the distinct perspectives adopted for deriving capacity. Traditionally, capacity is evaluated by calculating the number of patterns retrievable from the dynamics of neuronal firing activity. In contrast, our method focuses on the patterns directly stored within the distribution of synaptic weights.

Given that the information flow in the learning paradigm can be conceptualized as a Markov chain, where data are mapped to connectivity before being retrieved by dynamics, the data processing inequality \citep{cover1999elements} suggests that the mutual information between the data and connections must exceed that of the retrieval dynamics. Thus, our superlinear capacity result reflects the information loss inherent in the retrieval phase and serves as an upper bound for the practical capacity.

This echoes the perspective proposed in \citep{palm1996associative}, where associative memory is viewed as a communication channel consisting of two components: a storage stage, mapping data patterns to the synaptic connectivity matrix, and a retrieval stage, mapping the connectivity matrix to output patterns. These two components correspond to ``storage capacity" and ``retrieval capacity", respectively, each providing an upper bound on the total memory capacity when the full storage–retrieval procedure is considered. However, even with this perspective, related theoretical works \citep{palm1980associative, palm1992information} still define ``storage capacity" operationally as the amount of information about the patterns that can be successfully retrieved from the synaptic weights. This is achieved by applying the stored patterns to the learned connectivity matrix, performing thresholding detection, and quantifying the information content based on the retrieval error. Consequently, the resulting capacity is still derived from the perspective of neural activity, rather than the information content of the connectivity itself.

Separately, our approach provides a more general perspective on network coding. Traditional studies have focused primarily on the storage capacity of the \textit{full} $ d\times d $ Hopfield network, which limits their ability to examine ensembles consisting of only a subset of neurons or connections. By contrast, our method can be applied to \textit{any} selected ensemble of connections within the network, 
and reveals that the relationship between encoded information and synaptic ensemble size ($n$) follows $ MI \sim n^{1.1} $. This generalization, obtained by adopting a complementary perspective to the traditional one,  enables more detailed investigations of neural coding within different network substructures, such as by specific circuit motifs.

\subsection{Synergy among synaptic connections}

Our findings demonstrated that the information stored collectively by the connections in a network can exceed the sum of the information stored by individual synaptic connections; that is, $ \mathbb{E}[MI(\mathbf{w};\mathbf{x}^l)] > n \, \mathbb{E}[MI(w;\mathbf{x}^l)] $. This result followed directly from the positive correlation observed between $ \mathbb{E}[MI(\mathbf{w};\mathbf{x}^l)]/n $ and $ n $ (Fig. \ref{Fig:Synergy}B), specifically: $ \mathbb{E}[MI(\mathbf{w};\mathbf{x}^l)]/n > \mathbb{E}[MI(w;\mathbf{x}^l)]/1 $ when $ n > 1 $.

This phenomenon, where the whole is greater than the sum of its parts, is known as \textit{synergy} \citep{gawne1993independent, brenner2000synergy, williams2010nonnegative}. Information synergy has been a topic of long-standing interest in neuroscience, with evidence of its existence reported in ensembles of neurons \citep{panzeri1999correlations, schneidman2003synergy} and groups of brain regions \citep{luppi2022synergistic}. Our approach provides novel evidence of synergy at the level of synaptic connections.

Synergy in neural coding arises due to interactions among components of the system. These interactions can occur at various levels: pairwise, triplet, quadruplet, or even higher-order interactions. Classical studies of synergistic coding in neurons often focus on the role of correlations between spikes, which represent second-order interactions. More recent work has explored third-order interactions, introducing concepts such as triplet correlations \citep{cayco2015triplet}. By deriving closed-form expressions for Shannon mutual information, our work offers a straightforward way to analyze interactions at all levels—$2^{nd}$ order and higher—by calculating $ MI(\mathbf{w};\mathbf{x}^l)/n $, the contribution of a single connection to the encoding by the ensemble. In our simulations, we estimated interactions ranging from second to twentieth order using this value and observed a significant increasing trend as the number of connections $n$ grew (Fig. \ref{Fig:Synergy}BC). This suggests that synergistic effects extend at least to the 20th order, indicating that further insights may await discovery at higher orders. For instance, it remains unclear whether this synergistic scenario will continue indefinitely as interaction orders increase. Some studies suggest that synergistic effects dominate only at smaller neural populations, typically involving fewer than 200 neurons, while coding becomes increasingly redundant at larger scales \citep{kafashan2021scaling}; our analysis did not reach the scale of $n > 40,000$ synaptic connections due to computational limitations. Nonetheless, the linear regression results in Figure \ref{Fig:Synergy}C suggest a relationship of
$ MI \sim 0.001 n^2 $, which contrasts sharply with established results for Hopfield networks, providing further credence to our expectation that the linear trend seen in Figure \ref{Fig:Synergy}C will eventually plateau or saturate as $ n $ becomes large.

These observations leads to an intriguing question: what is the optimal size of an ensemble of neurons or synaptic connections to maximize synergy? Identifying this ``most synergistic" size may point to a fundamental unit of neural coding and represents a promising direction for future research.

\subsection{Potential implications for modern machine learning}

Here, we highlight several potential implications of our framework for the machine learning community.

First, we note that two adjacent layers in a multilayered deep neural network can be viewed as a dense bipartite network. Furthermore, several studies have explored progressively greater equivalence between the backpropagation algorithm and Hebbian-like local learning rules \citep{xie2003equivalence, harris2008stability, lillicrap2016random, akrout2019deep, fan2025contribute}. Under this interpretation, the data pattern $\mathbf{x}$ in our framework corresponds to the concatenation of the firing activities of two adjacent layers. The mutual information between the synaptic weights and the data patterns, $MI(\mathbf{w};\mathbf{x})$, thus measures the information encoded in the weights about the joint activity of the layers. This can be decomposed into two components: information about the preceding layer, and information that facilitates the prediction of the subsequent layer given the former. From this perspective, our framework provides a way to quantify how much information synaptic weights store about cross-layer co-activity. We speculate that the phenomena revealed by our framework, such as distributed coding and synergistic information, may also emerge between adjacent layers in modern deep neural networks.

Second, our work implies that pruning individual synapses based solely on their isolated information content may be misleading. Due to synergistic interactions, the contribution of a single synapse is inherently context-dependent and determined by the ensemble in which it is embedded. Pruning strategies that rely only on the mutual information between an individual synapse and the data may therefore underestimate its functional contribution. 

Finally, our framework enables comparisons of information storage across different network motifs that contain the same number of synapses but differ in connectivity patterns. This opens the door to future studies addressing which structural motifs play the most prominent roles in information storage. Understanding these relationships is a key step toward making ``black box" neural networks more interpretable and could potentially yield new neuroscientific insights.

\subsection{Assumptions and considerations}

In our exploration of the information encoded in synaptic ensembles, we chose to employ the densely connected neural network with Hebbian connectivity owing to its mathematical tractability.  
More commonly used architectures, such as multi-layer fully connected networks or recurrent neural networks, were not covered here. Also not considered were the additional complexities in biological neural networks, such 
as sparsity and spiking-timing-dependent-plasticity, beyond those captured by the Hebbian network model. Despite these constraints, Hebbian connectivity serves as a basis for various foundational models, such as Hopfield network, in associative memory and neural information storage. Indeed, it has been successfully applied to various brain regions, including the hippocampus \citep{sun2023organizing, kang2024distinguishing}, prefrontal cortex \citep{durstewitz2000neurocomputational}, primary visual cortex \citep{dong1992dynamic}, and olfactory bulb \citep{hopfield1991olfactory}, suggesting that insights from this model may apply more generally. Consequently, our framework represents an important step forward in characterizing information storage in synaptic connections. 

We also chose to model data patterns as being log-normally distributed. This assumption may appear artificial or restrictive at first glance. However, if the data patterns $\mathbf{x}$ in our framework were to be interpreted not as raw external inputs, but as the (internal) firing patterns evoked by those inputs at a preceding `layer' of neurons, this provides stable grounding for our assumption. This is because, whereas raw environmental data may take various forms, evoked neural firing rates are governed fundamentally by log-normal distribution across many parts of the nervous system \citep{buzsaki2014log}. Two mechanisms have previously been proposed to explain this phenomenon: First, the log-normal distribution could arise from the high nonlinearity of the neuronal transfer function \citep{roxin2011distribution}. Second, it could emerge from a linear transformation of log-normally distributed synaptic inputs \citep{wohrer2013population}. The latter is consistent with the heavy-tailed log-normal distribution of synaptic weights observed within the neocortex \citep{song2005highly, ikegaya2013interpyramid}, and potentially mirrors the log-normal structure of many natural stimuli, such as images \citep{huang1999statistics}, sounds \citep{voss19751}, and language \citep{piantadosi2014zipf}.

This said, our framework can, in fact, also be used to interpret information stored about raw external inputs, $\mathbf{x}_{\text{raw}}$. This is due to the invariance of Shannon mutual information under smooth, bijective transformations \citep{kraskov2004estimating}. As long as such a transformation exists between the raw stimulus $\mathbf{x}_{\text{raw}}$ and the resultant neural representation $\mathbf{x}$, the equality $I(\mathbf{w};\mathbf{x}_{\text{raw}}) = I(\mathbf{w};\mathbf{x})$ holds. Consequently, our theory can be understood to also describe how external, real-world patterns, such as images, are stored within Hebbian connectivities.

These characteristics suggest that insights from our framework go beyond the specific assumptions of the model, and may be broadly applicable to how synaptic connections encode information.

\subsection{Error analysis for Fenton-Wilkinson approximation}

When deriving the probability distribution for the weight ensemble, $p(\mathbf{w})$, we applied the Fenton-Wilkinson approximation to obtain a closed-form expression for subsequent derivations. This choice could introduce errors into the final mutual information estimates. Consequently, we investigated the mathematical properties of this error to provide practical insights into its control, and into heuristics regarding its dependency on other parameters. The complete mathematical derivations and proofs for the following claims are provided in the Supplementary Material.

Specifically, we prove that the error in $MI(\mathbf{w};\mathbf{x}^l)$ vanishes as the scale of a specific random variable $\epsilon$ approaches zero, provided that the ground-truth $\ln \mathbf{w}$ has a bounded probability density function with a finite $(2+\beta)^{\text{th}}$ moment for some positive real value $\beta > 0$, and that the deviation in the probability density estimation is an $L^1$ function.

Based on this analysis, the primary mechanism for managing the error is to limit the scale of $\epsilon$. Since the definition of $\epsilon$ itself is difficult to interpret, we identified a sufficient condition for controlling its magnitude: the variance of the log-transformed input data patterns should be small along their principal direction. In other words, ensuring that $\text{Var}_{\text{principal}}[\ln \mathbf{x}^k]$ is small for all patterns serves as a practical and interpretable way to suppress the approximation error.

We also derive heuristics regarding the dependency of the error on model parameters $d, n$, and $N$. By assuming that the error magnitude scales linearly with the scale of $\epsilon$, we conclude that the error is effectively independent of $d$ and $N$, while scaling as $\sqrt{n}$ with the size of the weight ensemble. Furthermore, higher correlations across the components of $\ln \mathbf{x}^k$ increase the error magnitude by intensifying the scale of $\epsilon$.

Finally, although our analysis primarily characterizes the limits under which the error vanishes rather than providing an exact quantitative bound, the practical utility of this analysis framework remains. Users can minimize errors by controlling $\text{Var}_{\text{principal}}[\ln \mathbf{x}^k]$ in practice. More importantly, the central conclusions of this work - distributed coding across weights, Hebbian capacity, and synergistic coding - are drawn from qualitative trends and slopes of the information curves. Therefore, these findings are robust to estimation errors and do not require perfectly precise numerical values to remain valid.

\section{Methods}

\subsection{Parameter setup for simulation}
In this section, we provide a detailed explanation of the three sets of simulations used in our analyses:

\noindent \textbf{First Simulation Set:} 
We generated $ N $ independent data patterns, $ \{\mathbf{x}^1, \dots, \mathbf{x}^N\} $,
where each data pattern $\mathbf{x}^k$ was a $d$-dimensional random variable following a multivariate log-normal distribution: $ \ln(\mathbf{x}^k) \sim \mathcal{N}(\mathbf{\mu}^k, \Sigma^k) $. The components of each $\mu^k$  were  sampled from a uniform distribution between -1 and 1, and the covariance matrix was modeled as an exchangeable matrix:
\begin{equation*}
\renewcommand{\arraystretch}{0.5} 
\setlength{\arraycolsep}{3pt} 
\Sigma^1 = \Sigma^2 = ... = \Sigma^N =
\begin{pmatrix}
\tilde{\sigma}^2 & \tilde{\rho} \tilde{\sigma}^2 & \hdots & \tilde{\rho}\tilde{\sigma}^2 \\
\tilde{\rho}\tilde{\sigma}^2 & \tilde{\sigma}^2 & \hdots &  \tilde{\rho}\tilde{\sigma}^2 \\
\vdots & \vdots & \ddots & \vdots \\
\tilde{\rho}\tilde{\sigma}^2 & \tilde{\rho}\tilde{\sigma}^2 & \hdots & \tilde{\sigma}^2
\end{pmatrix}.
\end{equation*}
The dimension $ d $ was chosen to be $20$.

We varied the parameters $N$, $n$, $\tilde{\sigma}$, and $\tilde{\rho}$ across different analyses. In Figure \ref{Fig:MI and pattern number}A, we set $n=1$, $\tilde{\rho} = 0 $, systematically varied $ N $ and $ \tilde{\sigma} $, and calculated the corresponding values of $ MI(\mathbf{x}^1; w_{23}) $. In Figure \ref{Fig:MI and pattern number}B, we set $n=1$, $ \tilde{\sigma} = 1 $, varied $ N $ and $ \tilde{\rho} $, and again calculated $ MI(\mathbf{x}^1; w_{23}) $. In Figure \ref{Fig:MI & ensemble size}A, we set $ N = 10 $, and $ \tilde{\rho} = 0 $, varied the size $ n $ by incrementally adding one connection from the set of randomly selected 15 connections to the synaptic ensemble $\mathbf{w}$, as well as varied $ \tilde{\sigma} $, and computed the corresponding values of $ MI(\mathbf{x}^1; \mathbf{w}) $. In Figure \ref{Fig:MI & ensemble size}B, we set $ N = 10 $, and $ \tilde{\sigma} = 1 $, varied $ n $ in the same way as above, as well as varied $ \tilde{\rho} $, and computed $ MI(\mathbf{x}^1; \mathbf{w}) $. In all cases where $ \tilde{\rho} $ was varied, it was restricted to the range $ \frac{-1}{d-1} \leq \tilde{\rho} \leq 1 $. This constraint arises from the requirement that the covariance matrices remain positive semi-definite. Specifically, the eigenvalues of the exchangeable covariance matrix $ \Sigma^k $ are $ \tilde{\sigma}^2 (1 - \tilde{\rho}) $ with multiplicity $ d - 1 $, and $ \tilde{\sigma}^2 (1 + (d - 1) \tilde{\rho}) $ with multiplicity 1; ensuring non-negativity for all eigenvalues imposes the stated range on $ \tilde{\rho} $.

\noindent \textbf{Second Simulation Set:} The objective of this simulation set was to sample multiple distribution-information ``tuples", i.e., $ \left(\{\mathbf{x}^1, \dots, \mathbf{x}^N\}, \mathbf{w}, MI(\mathbf{w}; \mathbf{x}^l)\right) $, and to analyze potential relationships between $ MI $ and distribution parameters based on these samples. This allowed for a more general exploration of the dependence of $ MI $ on the parameters than the one with the first simulation set, which only varied the values of specific individual parameters. To this end, in this set, instead of systematically varying parameters across ranges, all distribution parameters were sampled randomly. The ensemble size $n$ was randomly selected from 1 to 20 with equal probability. The dimension was set to $ d = 20 $, and the number of patterns $ N $ was randomly selected from 2 to 50 with equal probability. The parameters $ \{\mu^1, \dots, \mu^N\} $ were sampled from a uniform distribution between -1 and 1 for each component. The covariance matrices were identical for all patterns but more general than the previous set:
\begin{equation*}
\renewcommand{\arraystretch}{0.5} 
\setlength{\arraycolsep}{3pt} 
\Sigma^1 = \Sigma^2 = ... = \Sigma^N = \mathbf{A}^T\mathbf{A}/d +
\begin{pmatrix}
\tilde{\sigma}^2 & \tilde{\rho} \tilde{\sigma}^2 & \hdots & \tilde{\rho}\tilde{\sigma}^2 \\
\tilde{\rho}\tilde{\sigma}^2 & \tilde{\sigma}^2 & \hdots &  \tilde{\rho}\tilde{\sigma}^2 \\
\vdots & \vdots & \ddots & \vdots \\
\tilde{\rho}\tilde{\sigma}^2 & \tilde{\rho}\tilde{\sigma}^2 & \hdots & \tilde{\sigma}^2
\end{pmatrix}.
\end{equation*}
Here, $ \tilde{\sigma} $ was drawn from the absolute values of a standard normal distribution, and $ \tilde{\rho} $ was sampled from a uniform distribution in the range $ [\frac{-1}{d-1}, 1] $. The matrix $ \mathbf{A} $ is stochastic, with each element sampled from a standard normal distribution. Both components, $ \mathbf{A}^T \mathbf{A} / d $ and the exchangeable covariance matrix, are positive semi-definite, ensuring that their sum is a valid covariance matrix. The introduction of $ \mathbf{A}^T \mathbf{A} / d $ adds degrees of freedom, but it tends to produce matrices with larger diagonal elements, potentially neglecting distributions with highly correlated components. In contrast, the original exchangeable matrix can capture high correlations between components due to its structure, but it lacks sufficient degrees of freedom to account for more complex variability in the data. The $ 1/d $ scaling factor was designed to prevent either component from dominating.
Consequently, by combining the two components, the resulting covariance matrix strikes a balance between flexibly capturing a wider range of data patterns, 
and the ability to model correlations, thereby providing a richer and more balanced representation of the underlying data distributions.

Using this setup, we generated 1000 samples of $ \left(\{\mathbf{x}^1, \dots, \mathbf{x}^N\}, \mathbf{w}, MI(\mathbf{w}; \mathbf{x}^l)\right) $. For Figure \ref{Fig:MI and pattern number}C, we analyzed the regression between $ MI $ and $ N $. For Figure \ref{Fig:MI & ensemble size}C, we analyzed the regression between $ MI $ and $ n $.

\noindent \textbf{Third Simulation Set:} 
The third set of simulations followed a similar approach to the second in that we sampled numerous distribution-information ``tuples" $ \left(\{\mathbf{x}^1, \dots, \mathbf{x}^N\}, \mathbf{w}, MI(\mathbf{w}; \mathbf{x}^l) \right)$, the ensemble size $n$ was randomly selected from 1 to 20 with equal probability, the dimension $d$ was set to $20 $,  the number of patterns $ N $ was randomly selected from 2 to 50 with equal probability, and the parameters $ \{\mu^1, \dots, \mu^N\} $ were sampled from a uniform distribution between -1 and 1 for each component. The key difference was that the covariance matrices varied across patterns, following: 
\begin{equation*}
\Sigma^k  = \mathbf{Q}\mathbf{\Lambda}\mathbf{Q}^T + r \mathbf{1}\mathbf{1}^T.
\end{equation*}
Here, $ \mathbf{Q} $ is a random orthogonal matrix generated using the common practice of performing QR decomposition on a matrix whose elements were drawn from a standard normal distribution. $ \mathbf{\Lambda} $ is a diagonal matrix with random positive eigenvalues sampled from a normal distribution (mean 0.01, standard deviation 1.2), which is slightly broader than the standard normal to enable richer representations. The scalar factor $ r $ was sampled from a uniform distribution in the range [0, 1.44] to prevent either component from dominating.  Similar to the second simulation set, by combining these two components in the construction of the covariance matrices, we captured both the flexibility offered by $\mathbf{Q}\mathbf{\Lambda}\mathbf{Q}^T$ and the structure in $\mathbf{1}\mathbf{1}^T$ needed to model strong correlations across components. 

Using this setup, we generated 1000 samples. In Figure \ref{Fig:MI and pattern number}D, we analyzed the regression between $ MI $ and $ N $, while in Figure \ref{Fig:MI & ensemble size}D, we analyzed the regression between $ MI $ and $ n $. For the analysis of $ MI/n $ in Figure \ref{Fig:Synergy}, we generated an additional 20,000 samples.

\subsection{Proofs for propositions}
\label{sec: proofs}

\begin{proposition}{1.1}
\label{prop：pwij}
For $N$ independent log-normally distributed patterns $\{\mathbf{x}^1, ...,\mathbf{x}^N \}$,  the distribution of the synaptic connection $w_{ij}=\sum_{k=1}^N\mathbf{x}^k_i\mathbf{x}^k_j$ in the Hebbian network is approximated by a new log-normal distribution:
\begin{equation}
    p(w_{ij}) \approx \frac{1}{w_{ij}\sigma_{w_{ij}} \sqrt{2\pi}} \exp\left(
    -\frac{(\ln w_{ij} - \mu_{w_{ij}})^2}{2\sigma_{w_{ij}}^2}
    \right)
\end{equation}
with $\mu_{w_{ij}}, \sigma_{w_{ij}}$ defined in equation \ref{supp eq: mu_w}, \ref{supp eq: sigma_w}.
\end{proposition}

\begin{proof}
 Each data pattern is assumed to follow $\ln(\mathbf{x}^k) \sim \mathcal{N}(\mathbf{\mu}^k, \Sigma^k)$ in equation \ref{eq: pattern log normal}. Since the sum of two components in a multivariate normal distribution is normal, we have: 
\begin{equation}
 \ln(\mathbf{x}^k_i\mathbf{x}^k_j) = \ln(\mathbf{x}^k_i) + \ln(\mathbf{x}^k_j) \sim \mathcal{N}(\mathbf{\mu}^k_i + \mathbf{\mu}^k_j, (\sigma^k_i)^2 + (\sigma^k_j)^2 + 2\sigma^k_{ij}),
 \label{supp eq: lnxx~N}
\end{equation}
which means any term of the product form $\mathbf{x}^k_i\mathbf{x}^k_j$ follows a log-normal distribution. The sum of such terms from $N$ independent data patterns, $\sum_{k=1}^N \mathbf{x}^k_i \mathbf{x}^k_j$,  represents the connection weight between two neurons, $w_{ij}$, as defined as  in equation \ref{eq:hopWts}. Therefore, studying the distribution of $w_{ij}$ is equivalent to studying the distribution of the sum of several independent log-normal random variables. Using Fenton-Wilkinson method, we can approximate this distribution, $p(w_{ij})$, by a new log-normal distribution $p(\Hat w_{ij}) \approx p(w_{ij})$ with the same first and second moments. In other words, we aim to derive a log-normal distribution:
\begin{equation}
\ln(\Hat{w}_{ij}) \sim \mathcal{N}\left(\mu_{w_{ij}}, (\sigma_{w_{ij}})^2\right)
\label{supp eq: hatw_ij}
\end{equation}
whose first two moments match those of $w_{ij}$:
\begin{equation*}
\mathbb{E}[\Hat{w}_{ij}] = \mathbb{E}[w_{ij}], \; \mathrm{Var}[\Hat{w}_{ij}] =   \mathrm{Var}[w_{ij}].
\end{equation*}
The values for $\mathbb{E}[w_{ij}]$ and $\mathrm{Var}[w_{ij}]$ can be derived based on independence among patterns and the properties of the log-normal distribution\footnote{For random variable $\ln X \sim \mathcal{N}(\mu, \sigma^2)$, $\mathbb{E}[X] = e^{\mu + \frac{1}{2}\sigma^2}, \mathrm{Var}[X] = [e^{\sigma^2}-1]e^{2\mu + \sigma^2}$.}. For convenience, we denote them as $M_1$ and $M_2$ from here on. 
\begin{align}
    M_1 &:= \mathbb{E}[w_{ij}] = \mathbb{E}[\sum_{k=1}^N\mathbf{x}^k_i\mathbf{x}^k_j] 
    = \sum_{k=1}^N \mathbb{E}[\mathbf{x}^k_i\mathbf{x}^k_j] \nonumber\\
    &= \sum_{k=1}^N \exp\left(\mathbf{\mu}^k_i + \mathbf{\mu}^k_j + 
    \frac{1}{2}\left[(\sigma^k_{i})^2 + (\sigma^k_{j})^2 + 2\sigma^k_{ij}\right] \right) \label{supp eq: M1}\\
   M_2 &:= \mathrm{Var}[w_{ij}] = \mathrm{Var}[\sum_{k=1}^N\mathbf{x}^k_i\mathbf{x}^k_j] = \sum_{k=1}^N \mathrm{Var}[\mathbf{x}^k_i\mathbf{x}^k_j] \nonumber\\
    &= \sum_{k=1}^N \left[ \exp\left((\sigma^k_{i})^2 + (\sigma^k_{j})^2 + 2\sigma^k_{ij}\right) - 1 \right] \exp\left(2 (\mathbf{\mu}^k_i + \mathbf{\mu}^k_j) +
    (\sigma^k_{i})^2 + (\sigma^k_{j})^2 + 2\sigma^k_{ij} \right) \label{supp eq: M2}
\end{align}

Given $\mathbb{E}[\Hat{w}_{ij}]$ and $\mathrm{Var}[\Hat{w}_{ij}]$, the 
the parameters $\mu_{w_{ij}}$ and $\sigma^2_{w_{ij}}$ in equation \ref{supp eq: hatw_ij} are derived as follows:
\begin{align}
    \mu_{w_{ij}} &= \ln M_1 - \frac{1}{2}\ln(1 + \frac{M_2}{M_1^{\ 2}}),\label{supp eq: mu_w}\\
    \sigma_{w_{ij}}^2 &= \ln(1 + \frac{M_2}{M_1^{\ 2}}). \label{supp eq: sigma_w}
\end{align}
In this way, we obtain a distribution for a connection weight that is fully parametrized by the known data pattern distributions.
\end{proof}

\begin{proposition}{1.2}
The mutual information between a synaptic connection $w_{ij}$ and a data pattern $\mathbf{x}^l$ is given by the analytical approximation:
\begin{equation*}
    MI(w_{ij}; \mathbf{x}^l) = \left(\mu_{w_{ij}} - \mu_{w_{ij/l}}\right) + \left(\ln \sigma_{w_{ij}} - \ln \sigma_{w_{ij/l}}\right),
\end{equation*}
with $\mu_{w_{ij}}, \sigma_{w_{ij}}$ defined in equation \ref{supp eq: mu_w}, \ref{supp eq: sigma_w}; $\mu_{w_{ij/l}}, \sigma_{w_{ij/l}}$ defined in equation \ref{supp eq: mu_w/l}, \ref{supp eq: sigma_w/l}. 
The unit of $MI$ is nat.
\end{proposition}

\begin{proof}
We can calculate the mutual information $MI(w_{ij};\mathbf{x}^l)$ as the difference of differential entropies $h(w_{ij}) - h(w_{ij}|\mathbf{x}^l)$, which leads to:
\begin{align*}
    &MI(w_{ij};\mathbf{x}^l) = h(w_{ij}) - h(w_{ij}|\mathbf{x}^l) \\
    &= h(w_{ij})
    + \iint p(\mathbf{x}^l) p(w_{ij} |\mathbf{x}^l) \, \mathrm{log} \  p(w_{ij}|\mathbf{x}^l) \, \mathrm{d}w_{ij} \mathrm{d}\mathbf{x}^l \\
    &= h(w_{ij})
    + \int p(\mathbf{x}^l) \left( \int p(w_{ij} |\mathbf{x}^l) \, \mathrm{log} \  p(w_{ij}|\mathbf{x}^l) \, \mathrm{d}w_{ij} \right) \mathrm{d}\mathbf{x}^l \\
    &= h(w_{ij}) - \int p(\mathbf{x}^l) h(\sum_{k\neq l}^N\mathbf{x}^k_i\mathbf{x}^k_j) \mathrm{d}\mathbf{x}^l \\
    &= h(w_{ij}) - h(\sum_{k\neq l}^N\mathbf{x}^k_i\mathbf{x}^k_j)
\end{align*}
In the second-to-last step, we use the fact that $p(w_{ij}|\mathbf{x}^l) $ and $p(\sum_{k\neq l}^N\mathbf{x}^k_i\mathbf{x}^k_j )$ have the same differential entropy. This is because, given independence among patterns, $p(w_{ij}|\mathbf{x}^l):= p(\sum_{k}^N\mathbf{x}^k_i\mathbf{x}^k_j|\mathbf{x}^l) $ and $p(\sum_{k\neq l}^N\mathbf{x}^k_i\mathbf{x}^k_j )$ are shifted versions of each other. For convenience, we denote $w_{ij/l}:=\sum_{k\neq l}^N\mathbf{x}^k_i\mathbf{x}^k_j$ in what follows. Observe that the same reasoning represented in Proposition 1.1 also extends to $w_{ij/l}$, indicating it too can be also approximated by a log-normal distribution.

The $MI(w_{ij};\mathbf{x}^l)$ is therefore the difference between the differential entropies of two log-normal variables, $w_{ij}$ and $w_{ij/l}$. Considering the property of log-normal random variables\footnote{For random variable $\ln X \sim \mathcal{N}(\mu, \sigma^2)$, its entropy equals $\ln(\sqrt{2\pi}\sigma e^{\mu+\frac{1}{2}})$ in nats.}, we have:
\begin{align*}
   h(w_{ij}) &= \ln (\sqrt{2\pi}\sigma_{w_{ij}} e^{\mu_{w_{ij}}+\frac{1}{2}}), \\
   h(w_{ij/l}) &= \ln (\sqrt{2\pi}\sigma_{w_{ij/l}} e^{\mu_{w_{ij/l}}+\frac{1}{2}}).
\end{align*}

Consequently, we obtain the expression for mutual information: 
\begin{align*}
    MI(w_{ij};\mathbf{x}^l) &= \ln (\sqrt{2\pi}\sigma_{w_{ij}} e^{\mu_{w_{ij}}+\frac{1}{2}}) - \ln (\sqrt{2\pi}\sigma_{w_{ij/l}} e^{\mu_{w_{ij/l}}+\frac{1}{2}}) \\
    &= \left(\mu_{w_{ij}} - \mu_{w_{ij/l}}\right) + \left(\ln \sigma_{w_{ij}} - \ln \sigma_{w_{ij/l}} \right)
\end{align*}

For the sake of completeness, we explicitly write the parameters for $p(w_{ij/l})$ here. By adapting equation \ref{supp eq: hatw_ij} to \ref{supp eq: sigma_w}, we have:
\begin{equation*}
    \ln w_{ij/l}\overset{\text{approx.}}{\sim} \mathcal{N}\left(\mu_{w_{ij/l}}, (\sigma_{w_{ij/l}})^2\right),
\end{equation*}
where:
\begin{align}
    \mu_{w_{ij/l}} &= \ln M_{1/l} - \frac{1}{2}\ln(1 + \frac{M_{2/l}}{M_{1/l}^{\ 2}}) \label{supp eq: mu_w/l}\\
    \sigma_{w_{ij/l}}^2 &= \ln(1 + \frac{M_{2/l}}{M_{1/l}^{\ 2}}) \label{supp eq: sigma_w/l}\\
    M_{1/l} &= \sum_{k\neq l}^N \exp\left(\mathbf{\mu}^k_i + \mathbf{\mu}^k_j + 
    \frac{1}{2}\left[(\sigma^k_{i})^2 + (\sigma^k_{j})^2 + 2\sigma^k_{ij}\right] \right) \nonumber\\
    M_{2/l} &= \sum_{k=1}^N \left[ \exp\left(( \sigma^k_{i})^2 + (\sigma^k_{j})^2 + 2\sigma^k_{ij}\right) - 1 \right] \times \nonumber \\ 
    &\qquad \exp\left(2 (\mathbf{\mu}^k_i + \mathbf{\mu}^k_j) +
    (\sigma^k_{i})^2 + (\sigma^k_{j})^2 + 2\sigma^k_{ij} \right) \nonumber
\end{align}

\end{proof}

\begin{proposition}{2.1}
For a pair of synaptic connections $\mathbf{w}=(w_{ij}, w_{mn})^T$ following a 2-dimensional log-normal distribution, the covariance between $\ln w_{ij}$ and $\ln w_{mn}$ is:
\begin{align*}
&\sigma_{w_{ij,mn}} = \ln ( 1 \; + \nonumber \\ 
&\frac{\sum_{k=1}^N \exp\left(
\sum_{\phi}^{\{i,j,m,n\}} \mu^k_{\phi} 
+ \frac{1}{2}\sum_{\phi}^{\{i,j,m,n\}} (\sigma^k_{\phi})^2
+ \sum_{\phi}^{\{ij,mn\}} \sigma^k_{\phi}
\right)\left[\exp\left( 
\sum_{\phi}^{\{i,j\}\times\{m,n\}} \sigma^k_{\phi}
\right) - 1\right]}
{\sum_{k,l=1}^N \exp\left( 
\sum_{\phi}^{\{i,j\}} \mu^k_{\phi} 
+ \sum_{\phi}^{\{m,n\}} \mu^l_{\phi} 
+ \frac{1}{2}\sum_{\phi}^{\{i,j\}} (\sigma^k_{\phi})^2
+ \frac{1}{2}\sum_{\phi}^{\{m,n\}} (\sigma^l_{\phi})^2
+ \sigma_{ij}^k + \sigma_{mn}^l
\right)}
).
\end{align*}
\end{proposition}

\begin{proof}
We have assumed that the ensemble $\mathbf{w}$ approximately follows 2-dimensional log-normal distribution, which states:
\begin{equation*}
\renewcommand{\arraystretch}{0.6} 
\setlength{\arraycolsep}{3pt} 
\ln \mathbf{w} = 
\begin{pmatrix}
\ln w_{ij} \\
\ln w_{mn}
\end{pmatrix} \overset{\text{approx.}}{\sim} \mathcal{N} (
\mu_{\mathbf{w}} = 
\begin{pmatrix}
\mu_{w_{ij}} \\
\mu_{w_{mn}}
\end{pmatrix}, 
\Sigma_\mathbf{w} =
\begin{pmatrix}
\sigma_{w_{ij}}^2 & \sigma_{w_{ij, mn}} \\
\sigma_{w_{mn,ij}} & \sigma_{w_{mn}}^2
\end{pmatrix}
).
\end{equation*}
Given the property of a multivariate log-normal random variable\footnote{For multivariate random variable $\ln X \sim \mathcal{N}(\mu, \Sigma), \mathrm{Var}[X]_{ij}=(e^{\Sigma_{ij}}-1)e^{\mu_i + \mu_j + \frac{1}{2}(\Sigma_{ii} + \Sigma_{jj})}$}, the covariance between two components of random variable $\mathbf{w}$, $\mathrm{Cov}[w_{ij}, w_{mn}]$, and the covariance between their logarithms, $\sigma_{w_{ij,mn}}$, are related by the following equation:
\begin{equation*}
    \mathrm{Cov}[w_{ij}, w_{mn}] = \left[\exp(\sigma_{w_{ij,mn}}) -1 \right]
    \exp\left(\mu_{w_{ij}} + \mu_{w_{mn}}+\frac{1}{2}(\sigma_{w_{ij}}^2 + \sigma_{w_{mn}}^2)\right),
\end{equation*}
which is equivalent to:
\begin{equation}
    \sigma_{w_{ij,mn}} =\ln \left(1 + \frac{\mathrm{Cov}[w_{ij}, w_{mn}]}{\exp\left[\mu_{w_{ij}} + \mu_{w_{mn}}+\frac{1}{2}(\sigma_{w_{ij}}^2 + \sigma_{w_{mn}}^2)\right]}\right) \label{supp eq: sigma=f(Cov[ww])}
\end{equation}

Given that the expressions for $\mu_{w_{ij}}, \mu_{w_{mn}}$ and $\sigma_{w_{ij}}^2,\sigma_{w_{mn}}^2$ have been derived in equation \ref{supp eq: mu_w} and \ref{supp eq: sigma_w}, the problem reduces to finding  $\mathrm{Cov}[w_{ij}, w_{mn}]$. From its definition, we have:
\begin{align}
    &\mathrm{Cov}[w_{ij}, w_{mn}] = \mathbb{E}[w_{ij} w_{mn}] - \mathbb{E}[w_{ij}]\mathbb{E}[w_{mn}] \nonumber \\
    &= \mathbb{E} \left[ (\sum_{k=1}^N \mathbf{x}^k_i\mathbf{x}^k_j) (\sum_{l=1}^N \mathbf{x}^l_m\mathbf{x}^l_n)\right] - 
    \mathbb{E} \left[ \sum_{k=1}^N \mathbf{x}^k_i\mathbf{x}^k_j \right] \mathbb{E}\left[ \sum_{l=1}^N \mathbf{x}^l_m\mathbf{x}^l_n\right] \nonumber \\
    &=   \sum_{k,l=1}^N \mathbb{E} \left[\mathbf{x}^k_i\mathbf{x}^k_j  \mathbf{x}^l_m\mathbf{x}^l_n\right] - 
     \sum_{k=1}^N \mathbb{E} \left[\mathbf{x}^k_i\mathbf{x}^k_j \right]  \sum_{l=1}^N \mathbb{E}\left[\mathbf{x}^l_i\mathbf{x}^l_j\right] \nonumber\\
     &=  \left( \sum_{k\neq l \,  \, k,l=1}^N \mathbb{E} \left[\mathbf{x}^k_i\mathbf{x}^k_j  \mathbf{x}^l_m\mathbf{x}^l_n\right]
     + \sum_{k=1}^N \mathbb{E} \left[\mathbf{x}^k_i\mathbf{x}^k_j  \mathbf{x}^k_m\mathbf{x}^k_n\right] \right)
     - \sum_{k,l=1}^N \mathbb{E} \left[\mathbf{x}^k_i\mathbf{x}^k_j \right]  \mathbb{E}\left[\mathbf{x}^l_i\mathbf{x}^l_j\right] \nonumber\\
     &=   \sum_{k\neq l \,  \, k,l=1}^N \mathbb{E} \left[\mathbf{x}^k_i\mathbf{x}^k_j  \right]  \mathbb{E}\left[\mathbf{x}^l_m\mathbf{x}^l_n\right]
     + \sum_{k=1}^N \mathbb{E} \left[\mathbf{x}^k_i\mathbf{x}^k_j  \mathbf{x}^k_m\mathbf{x}^k_n\right]
     - \sum_{k,l=1}^N \mathbb{E} \left[\mathbf{x}^k_i\mathbf{x}^k_j \right]  \mathbb{E}\left[\mathbf{x}^l_i\mathbf{x}^l_j\right] \nonumber \\
     &= \sum_{k=1}^N \left( \mathbb{E} \left[\mathbf{x}^k_i\mathbf{x}^k_j  \mathbf{x}^k_m\mathbf{x}^k_n\right] - \mathbb{E} \left[\mathbf{x}^k_i\mathbf{x}^k_j \right]
     \mathbb{E}\left[\mathbf{x}^k_m\mathbf{x}^k_n\right] \right). \label{supp eq: Exxxx-ExEx}
\end{align}
The second-to-last step uses the independence condition between data different patterns. 

To derive the term $\mathbb{E} \left[\mathbf{x}^k_i\mathbf{x}^k_j  \mathbf{x}^k_m\mathbf{x}^k_n\right]$ in the last line, recall that $\mathbf{x}^k_i, \mathbf{x}^k_j,  \mathbf{x}^k_m, \mathbf{x}^k_n$ are four components of the same multivariate log-normal random vector $\mathbf{x}^k$. Thus, the product $\mathbf{x}^k_i\mathbf{x}^k_j \mathbf{x}^k_m\mathbf{x}^k_n$ is log-normally distributed. Therefore,  by extending the same reasoning in equation \ref{supp eq: lnxx~N}, we get:
\begin{align*}
\ln(\mathbf{x}^k_i\mathbf{x}^k_j \mathbf{x}^k_m\mathbf{x}^k_n) \sim \mathcal{N}
\left(\sum_{\phi}^{\{i,j,m,n\}} \mu^k_{\phi} \, ,
\sum_{\phi}^{\{i,j,m,n\}} (\sigma^k_{\phi})^2
+ \sum_{\phi}^{\{ij,mn\}} 2\sigma^k_{\phi}
+ \sum_{\phi}^{\{i,j\}\times\{m,n\}} 2\sigma^k_{\phi}
\right),
\end{align*}
where the different summations are defined as follows:
\begin{align*}
&\sum_{\phi}^{\{i,j,m,n\}} \mu^k_{\phi} = \mu^k_i + \mu^k_j + \mu^k_m + \mu^k_n ,\\
&\sum_{\phi}^{\{i,j,m,n\}} (\sigma^k_{\phi})^2 =(\sigma^k_i)^2 + (\sigma^k_j)^2 + (\sigma^k_m)^2 + (\sigma^k_n)^2 ,\\
&\sum_{\phi}^{\{ij,mn\}} 2\sigma^k_{\phi} = 2\sigma^k_{ij} +  2\sigma^k_{mn} \; ,
\sum_{\phi}^{\{i,j\}\times\{m,n\}} 2\sigma^k_{\phi} = 2\sigma^k_{im} + 2\sigma^k_{in} + 2\sigma^k_{jm} + 2\sigma^k_{jn}.
\end{align*}
Consequently, the mean of this quadruple product term is:
\begin{equation}
    \mathbb{E} \left[\mathbf{x}^k_i\mathbf{x}^k_j  \mathbf{x}^k_m\mathbf{x}^k_n\right] = \exp\left(
    \sum_{\phi}^{\{i,j,m,n\}} \mu^k_{\phi} 
    + \frac{1}{2}\sum_{\phi}^{\{i,j,m,n\}} (\sigma^k_{\phi})^2
    + \sum_{\phi}^{\{ij,mn\}} \sigma^k_{\phi}
    + \sum_{\phi}^{\{i,j\}\times\{m,n\}} \sigma^k_{\phi}
    \right). \label{supp eq: Exxxx}
\end{equation}

For the term $\mathbb{E} \left[\mathbf{x}^k_i\mathbf{x}^k_j \right]\mathbb{E}\left[\mathbf{x}^k_m\mathbf{x}^k_n\right]$ in the equation \ref{supp eq: Exxxx-ExEx}, given that $\mathbb{E} \left[\mathbf{x}^k_i\mathbf{x}^k_j \right]$ and $ \mathbb{E}\left[\mathbf{x}^k_m\mathbf{x}^k_n\right]$ were obtained as components in the summation in equation \ref{supp eq: M1}, we get:
\begin{equation}
    \mathbb{E} \left[\mathbf{x}^k_i\mathbf{x}^k_j \right]\mathbb{E}\left[\mathbf{x}^k_m\mathbf{x}^k_n\right] = \exp\left(
    \sum_{\phi}^{\{i,j,m,n\}} \mu^k_{\phi} 
    + \frac{1}{2}\sum_{\phi}^{\{i,j,m,n\}} (\sigma^k_{\phi})^2
    + \sum_{\phi}^{\{ij,mn\}} \sigma^k_{\phi}
    \right).\label{supp eq: ExxExx}
\end{equation}

Substituting the results from equations \ref{supp eq: Exxxx} and \ref{supp eq: ExxExx} into equation \ref{supp eq: Exxxx-ExEx}, we obtain the covariance between two synaptic weights:
\begin{align}
    \mathrm{Cov}[w_{ij}, w_{mn}] =&
    \sum_{k=1}^N \exp\left(
    \sum_{\phi}^{\{i,j,m,n\}} \mu^k_{\phi} 
    + \frac{1}{2}\sum_{\phi}^{\{i,j,m,n\}} (\sigma^k_{\phi})^2
    + \sum_{\phi}^{\{ij,mn\}} \sigma^k_{\phi}
    \right) \times \nonumber \\
    &\left[\exp\left( 
    \sum_{\phi}^{\{i,j\}\times\{m,n\}} \sigma^k_{\phi}
    \right) - 1\right]. \label{supp eq: Cov[ww]}
\end{align}

Finally, by combining the covariance $\mathrm{Cov}[w_{ij}, w_{mn}]$ with equations \ref{supp eq: mu_w}, \ref{supp eq: sigma_w}, \ref{supp eq: sigma=f(Cov[ww])}, we obtain an expression for $\sigma_{w_{ij,mn}}$ as a function of the parameters characterizing the data pattern distributions:
\begin{align}
&\sigma_{w_{ij,mn}} = \ln ( 1 \; + \nonumber\\ 
&\frac{\sum_{k=1}^N \exp\left(
\sum_{\phi}^{\{i,j,m,n\}} \mu^k_{\phi} 
+ \frac{1}{2}\sum_{\phi}^{\{i,j,m,n\}} (\sigma^k_{\phi})^2
+ \sum_{\phi}^{\{ij,mn\}} \sigma^k_{\phi}
\right)\left[\exp\left( 
\sum_{\phi}^{\{i,j\}\times\{m,n\}} \sigma^k_{\phi}
\right) - 1\right]}
{\sum_{k,l=1}^N \exp\left( 
\sum_{\phi}^{\{i,j\}} \mu^k_{\phi} 
+ \sum_{\phi}^{\{m,n\}} \mu^l_{\phi} 
+ \frac{1}{2}\sum_{\phi}^{\{i,j\}} (\sigma^k_{\phi})^2
+ \frac{1}{2}\sum_{\phi}^{\{m,n\}} (\sigma^l_{\phi})^2
+ \sigma_{ij}^k + \sigma_{mn}^l
\right)}
). \label{supp eq: sigma_wijmn}
\end{align}

\end{proof}

\begin{proposition}{2.2}
 Under the approximation that both $ \mathbf{w} := (w_{ij}, w_{mn})^T $ and $\mathbf{w}_{/l}:= (w_{ij/l}, w_{mn/l})^T =(\sum_{k\neq l}^N \mathbf{x}^k_i\mathbf{x}^k_j , \; \sum_{k\neq l}^N \mathbf{x}^k_m\mathbf{x}^k_n )^T$ follow a multivariate log-normal distribution, i.e. $\ln \mathbf{w} \sim \mathcal{N}(\mu_\mathbf{w}, \Sigma_\mathbf{w})$ and $\ln \mathbf{w}_{/l} \sim \mathcal{N}(\mu_{\mathbf{w}_{/l}}, \Sigma_{\mathbf{w}_{/l}})$, the mutual information between the pair of connections $\mathbf{w}$ and a data pattern $ \mathbf{x}^l $ can be expressed as:
\begin{equation*}
MI(\mathbf{w}; \mathbf{x}^l) = (\mu_{w_{ij}} + \mu_{w_{mn}} - \mu_{w_{ij/l}} - \mu_{w_{mn/l}}) + \frac{1}{2}\ln|\Sigma_\mathbf{w}| - \frac{1}{2}\ln|\Sigma_{\mathbf{w}_{/l}}|,
\end{equation*}
where the expressions for $\mu_{w_{ij}}$, $\mu_{w_{mn}}$ are given in equation \ref{supp eq: mu_w}, and for $\mu_{w_{ij/l}}$, $\mu_{w_{mn/l}}$ in equation \ref{supp eq: mu_w/l}. For matrix $\Sigma_{\mathbf{w}}$, the diagonal elements are given by equation \ref{supp eq: sigma_w}, and the off-diagonal elements by equation \ref{supp eq: sigma_wijmn}; for  $\Sigma_{\mathbf{w}_{/l}}$, the corresponding expressions are provided by equations \ref{supp eq: sigma_w/l} and  \ref{supp eq: sigma_wijmn/l}. Here, $ |\Sigma_\mathbf{w}| $ denotes the determinant of the matrix $ \Sigma_\mathbf{w} $. All information values are expressed in nats. 
\end{proposition}

\begin{proof}
Given that $\mathbf{w}$ follows a 2-dimensional log-normal variable, its differential entropy (in nats) can be written as\footnote{For $n$-dimensional multivariate random variable $\ln X \sim \mathcal{N}(\mu, \Sigma)$, its differential entropy $h(X) = \frac{1}{2}\ln((2\pi e)^n |\Sigma|) + \sum_i \mu_i$ in nats.}:  
\begin{equation*}
h(\mathbf{w}) = \frac{1}{2}\ln\left( (2\pi e)^2 |\Sigma_\mathbf{w}|\right) + \mu_{w_{ij}} + \mu_{w_{mn}}.
\end{equation*}

For $h(\mathbf{w}|\mathbf{x}^l)$, observe that $p(\mathbf{w}|\mathbf{x}^l)$ is a shifted version of $p(\mathbf{w}_{/l}) :=p(\sum_{k\neq l}^N \mathbf{x}^k_i\mathbf{x}^k_j , \; \sum_{k\neq l}^N \mathbf{x}^k_m\mathbf{x}^k_n )$ . Therefore, the two distributions have the same entropy. This leads to:
\begin{align*}
h(\mathbf{w}|\mathbf{x}^l) = h(\mathbf{w}_{/l})= \frac{1}{2}\ln\left( (2\pi e)^2 |\Sigma_{\mathbf{w}_{/l}}|\right) + \mu_{w_{ij/l}} + \mu_{w_{mn/l}}. 
\end{align*}
The expressions for $\mu_{w_{ij/l}}, \mu_{w_{mn/l}},$  and for the diagonal elements of the covariance matrix, $ \sigma_{w_{ij/l}}^2, \sigma_{w_{mn/l}}^2$, are given in the proof of Proposition 1.2. For the off-diagonal element $\sigma_{w_{ij, mn}/l}$, its derivation mirrors that of $\sigma_{w_{ij, mn}}$ in Proposition 2.1. For clarity, we present the explicit formula here:
\begin{align}
&\sigma_{w_{ij,mn/l}} = \ln ( 1 \; + \nonumber\\ 
&\frac{\sum_{k\neq l}^N \exp\left(
\sum_{\phi}^{\{i,j,m,n\}} \mu^k_{\phi} 
+ \frac{1}{2}\sum_{\phi}^{\{i,j,m,n\}} (\sigma^k_{\phi})^2
+ \sum_{\phi}^{\{ij,mn\}} \sigma^k_{\phi}
\right)\left[\exp\left( 
\sum_{\phi}^{\{i,j\}\times\{m,n\}} \sigma^k_{\phi}
\right) - 1\right]}
{\sum_{k,g\neq l}^N \exp\left( 
\sum_{\phi}^{\{i,j\}} \mu^k_{\phi} 
+ \sum_{\phi}^{\{m,n\}} \mu^g_{\phi} 
+ \frac{1}{2}\sum_{\phi}^{\{i,j\}} (\sigma^k_{\phi})^2
+ \frac{1}{2}\sum_{\phi}^{\{m,n\}} (\sigma^g_{\phi})^2
+ \sigma_{ij}^k + \sigma_{mn}^g
\right)}
).\label{supp eq: sigma_wijmn/l}
\end{align} 

With $h(\mathbf{w})$ and $h(\mathbf{w}|\mathbf{x}^l)$ computed, we can now obtain the mutual information between any pair of synaptic interconnections and a data pattern:
\begin{equation*}
    MI(\mathbf{w};\mathbf{x}^l) = h(\mathbf{w})- h(\mathbf{w}|\mathbf{x}^l) =(\mu_{w_{ij}} + \mu_{w_{mn}} - \mu_{w_{ij/l}} - \mu_{w_{mn/l}}) + \frac{1}{2}\ln|\Sigma_\mathbf{w}| - \frac{1}{2}\ln|\Sigma_{\mathbf{w}_{/l}}|.
\end{equation*}
\end{proof}

\begin{proposition}{3}
For a synaptic ensemble of $n$ connections, $\mathbf{w}:=(w_{ij(1)}, w_{ij(2)}, ..., w_{ij(n)})^T$, where each $w_{ij(k)}$ denotes the connection between neuron $i(k)$ and $j(k)$, and under the approximation that both $\mathbf{w}$ and $\mathbf{w}_{/l}:= (w_{ij(1)/l}, w_{ij(2)/l}, ..., w_{ij(n)/l})^T$ follow a multivariate log-normal distribution, i.e. $\ln \mathbf{w} \sim \mathcal{N}(\mu_\mathbf{w}, \Sigma_\mathbf{w})$ and $\ln \mathbf{w}_{/l} \sim \mathcal{N}(\mu_{\mathbf{w}_{/l}}, \Sigma_{\mathbf{w}_{/l}})$,
the mutual information between the joint activity of these $n$ synaptic connections $\mathbf{w}$ and a data pattern $\mathbf{x}^l$ is given by the analytical approximation: 
\begin{equation*}
     MI(\mathbf{w};\mathbf{x}^l) = \sum_{k=1}^n \left(\mu_{w_{ij(k)}}  - \mu_{w_{ij(k)/l}}\right) + \frac{1}{2}\ln|\Sigma_\mathbf{w}| - \frac{1}{2}\ln|\Sigma_{\mathbf{w}_{/l}}|.
\end{equation*}
where the expressions for $\mu_{w_{ij(k)}}$ are given in equation \ref{supp eq: mu_w}, and for $\mu_{w_{ij(k)/l}}$ in equation \ref{supp eq: mu_w/l}. The diagonal and off-diagonal elements of $\Sigma_{\mathbf{w}}$ are given by equations \ref{supp eq: sigma_w} and \ref{supp eq: sigma_wijmn}, respectively; for  $\Sigma_{\mathbf{w}_{/l}}$, by equations \ref{supp eq: sigma_w/l} and  \ref{supp eq: sigma_wijmn/l}, respectively. All information values are expressed in nats.
\end{proposition}

\begin{proof}
Based on the properties of an $n$-dimensional log-normal variable, the differential entropy of $\mathbf{w}$ is given by: 
\begin{equation*}
h(\mathbf{w}) = \frac{1}{2}\ln\left( (2\pi e)^n |\Sigma_\mathbf{w}|\right) + \sum_{k=1}^n \mu_{w_{ij(k)}}.
\end{equation*}
Meanwhile, $p(\mathbf{w}|\mathbf{x}^l)$ and $p(\mathbf{w}_{/l})$ have the same entropy owing to their shifting relation. Therefore, we have:
\begin{equation*}
h(\mathbf{w}|\mathbf{x}^l) = h(\mathbf{w}_{/l}) =\frac{1}{2}\ln\left( (2\pi e)^n |\Sigma_{\mathbf{w}_{/l}}|\right) + \sum_{k=1}^n \mu_{w_{ij(k)/l}}.
\end{equation*}
Consequently, the information stored in ensemble $\mathbf{w}$ about pattern $\mathbf{x}^l$ is: 
\begin{equation*}
    MI(\mathbf{w};\mathbf{x}^l) = h(\mathbf{w}) - h(\mathbf{w}|\mathbf{x}^l) = \sum_{k=1}^n \left(\mu_{w_{ij(k)}}  - \mu_{w_{ij(k)/l}}\right) + \frac{1}{2}\ln|\Sigma_\mathbf{w}| - \frac{1}{2}\ln|\Sigma_{\mathbf{w}_{/l}}|.
\end{equation*}
Since the derivation in Proposition 1.1, 1.2 is valid for any single connection $w_{ij(k)}$, and that in Proposition 2.1 for any pair, the expressions derived earlier can be directly applied to the quantities in this equation.
\end{proof}

\section*{Acknowledgments}
This work was supported by funding from a JHU Discovery Award co-funded with the One Neuro Initiative (SPM), and from NIH grant 2R01EY027718 (SPM).

\section*{Competing Interests}
The author declare no competing interests.

\section*{Author Contributions}
Conceptualization: X.F. and S.P.M.; Theoretical construction: X.F.; Formal analysis: X.F.; Funding acquisition: S.P.M.; Supervision: S.P.M.; Visualization: X.F. and S.P.M.; Writing and Editing: X.F. and S.P.M.

\section*{Code Availability}
The code used to generate the simulations and figures in this paper can be found on Github at https://github.com/PixelsForest/weight-information .

\bibliographystyle{apalike}
\bibliography{references} 

\newpage
\includepdf[pages=-]{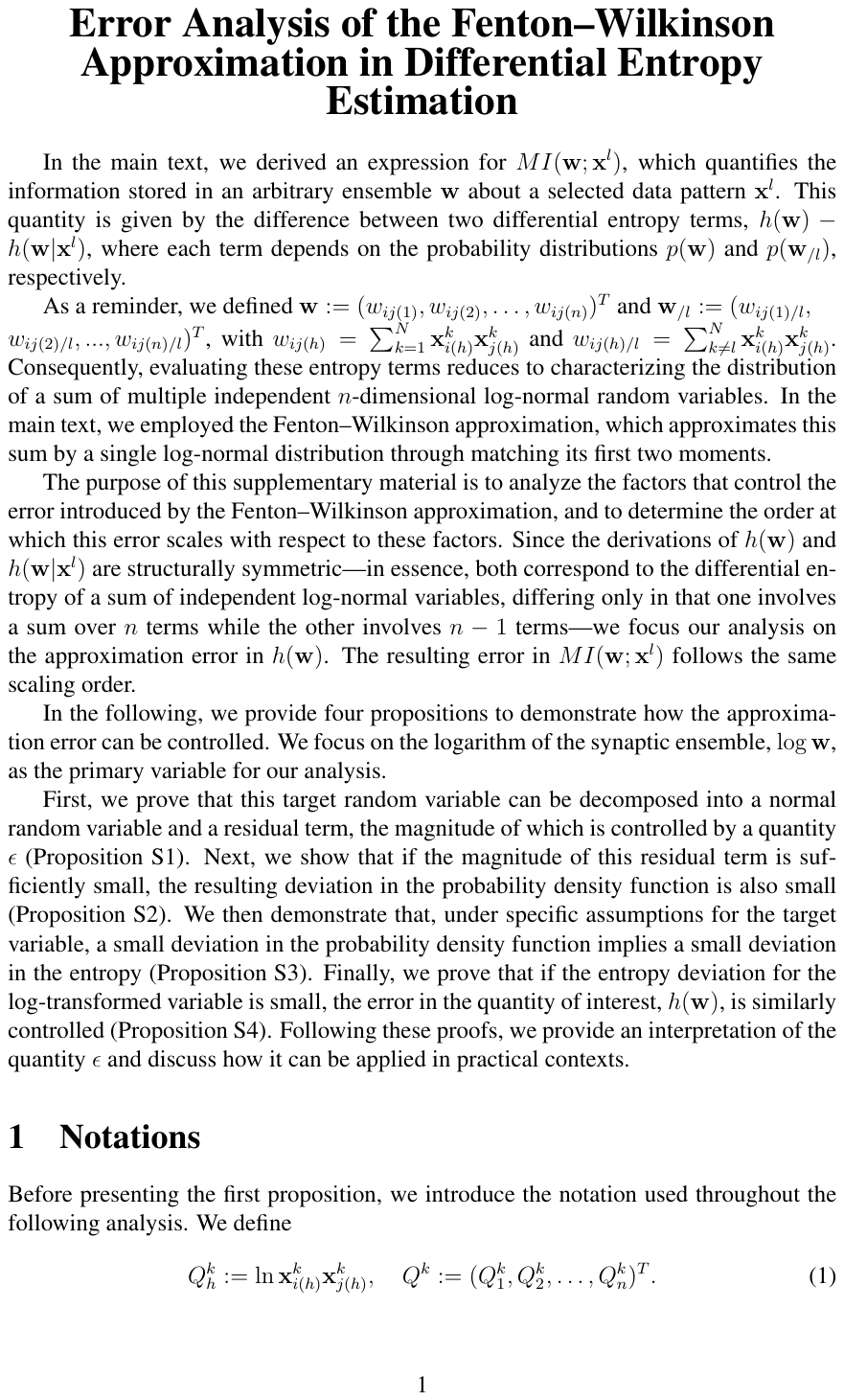}

\end{document}